\documentclass[prl,reprint,noeprint,showpacs,superscriptaddress]{revtex4-1}

\usepackage{graphicx}
\usepackage{url}
\usepackage{amsmath}
\usepackage{amssymb}
\usepackage{xfrac}
\usepackage{dsfont}
\usepackage{braket}
\usepackage{bm}
\usepackage{bbm}
\usepackage{color}

\newcommand{\affcua}{MIT-Harvard Center for Ultracold Atoms, Research Laboratory of Electronics, and Department of Physics,
Massachusetts Institute of Technology, Cambridge, Massachusetts 02139, USA}
\newcommand{\affens}{D\'{e}partement de Physique, Ecole Normale Sup\'{e}rieure / PSL Research University, CNRS, 24 rue Lhomond, 75005 Paris, France}

\begin{document}

\title{Boiling a Unitary Fermi Liquid}

\author{Zhenjie Yan}
\affiliation{\affcua}
\author{Parth B. Patel}
\affiliation{\affcua}
\author{Biswaroop Mukherjee}
\affiliation{\affcua}

\author{Richard J. Fletcher}
\affiliation{\affcua}

\author{Julian Struck}
\affiliation{\affcua}
\affiliation{\affens}

\author{Martin W. Zwierlein}
\affiliation{\affcua}

\begin{abstract}
We study the thermal evolution of a highly spin-imbalanced, homogeneous Fermi gas with unitarity limited interactions, from a Fermi liquid of polarons at low temperatures to a classical Boltzmann gas at high temperatures. Radio-frequency spectroscopy gives access to the energy, lifetime, and short-range correlations of Fermi polarons at low temperatures $T$. In this regime, we observe a characteristic $T^2$ dependence of the spectral width, corresponding to the quasiparticle decay rate expected for a Fermi liquid. At high $T$, the spectral width decreases again towards the scattering rate of the classical, unitary Boltzmann gas, $\propto T^{-1/2}$. In the transition region between the quantum degenerate and classical regime, the spectral width attains its maximum, on the scale of the Fermi energy, indicating the breakdown of a quasiparticle description.
Density measurements in a harmonic trap directly reveal the majority dressing cloud surrounding the minority spins and yield the compressibility along with the effective mass of Fermi polarons.
\end{abstract}


\maketitle

Landau's Fermi liquid theory provides a quasiparticle description of the low-temperature behavior for a large class of unordered fermionic states of matter, including most normal metals, atomic nuclei, and liquid \mbox{helium-3}~\cite{Nozieres1966}. Strongly interacting Fermi gases with highly imbalanced spin populations have been identified as belonging to the same class~\cite{Zwierlein2006d,Chevy2006a,Lobo2006d,Combescot2007b,Shin2008,Bruun2008,Prokofev2008a,Veillette2008,Schirotzek2009, Nascimbene2009, Nascimbene2010,Navon2010,Sommer2011b}. The quasiparticles in spin-imbalanced Fermi gases are Fermi polarons: spin impurities dressed by an excess cloud of majority fermions.
The stability of quasiparticles in a Fermi liquid is a consequence of the restricted phase space for collisions due to Pauli blocking. With increasing temperature $T$, the accessible phase space increases, and the lifetime of quasiparticles shortens, leading to the breakdown of Fermi liquid theory. In this intermediate temperature regime the gas is neither a Fermi liquid nor a classical Boltzmann gas. For strong interactions, this regime is void of well-defined quasiparticles and controlled by the quantum critical point of the unitary, spin-balanced gas at zero chemical potential and temperature~\cite{Nikolic2007,Enss2012,Frank2018}.

Ultracold Fermi gases offer a unique opportunity to study the crossover from a low-temperature Fermi liquid to a classical Boltzmann gas, due to the large accessible temperature range.
In spin-imbalanced Fermi gases, the two inequivalent Fermi surfaces provide additional richness. As the temperature is lowered from the classical regime, the Fermi surface of the majority forms first, giving minority spins the quasiparticle character of polarons. At even lower temperatures, the polarons themselves become quantum degenerate and form a Fermi surface.

In this work, we access the entire crossover from degenerate polarons to the classical Boltzmann gas through the quantum critical region. The internal properties of the polaronic quasiparticles are measured via radio-frequency (rf) spectroscopy~\cite{Schirotzek2009,Kohstall2012a,Koschorreck2012,Scazza2017} on a homogeneous Fermi gas~\cite{Mukherjee2017b,Hueck2018}. At low temperatures, the peak position and width of the rf spectra reflect energy and decay rate of the polarons. Note that the decay rate of a quasiparticle can be viewed as the rate of momentum relaxation in a transport measurement (see, e.g., \cite{Bruun2008}). The wings of the rf spectra yield information about the short-range correlations and the contact~\cite{Tan2008,Baym2007,Punk2007,Schneider2009,Braaten2010}, controlling the change in the polaron energy with interaction strength. Further thermodynamic properties of the polaron gas are directly obtained from \emph{in situ} density profiles in the presence of a harmonic potential~\cite{Shin2008a,Shin2008,Nascimbene2010,Navon2010,Horikoshi2010,Ku2012}, revealing the number of atoms in the majority dressing cloud of a polaron. The compressibility of the impurity gas at low temperature yields the effective mass of Fermi polarons.



\begin{figure*}
\centering
\includegraphics[width=\textwidth]{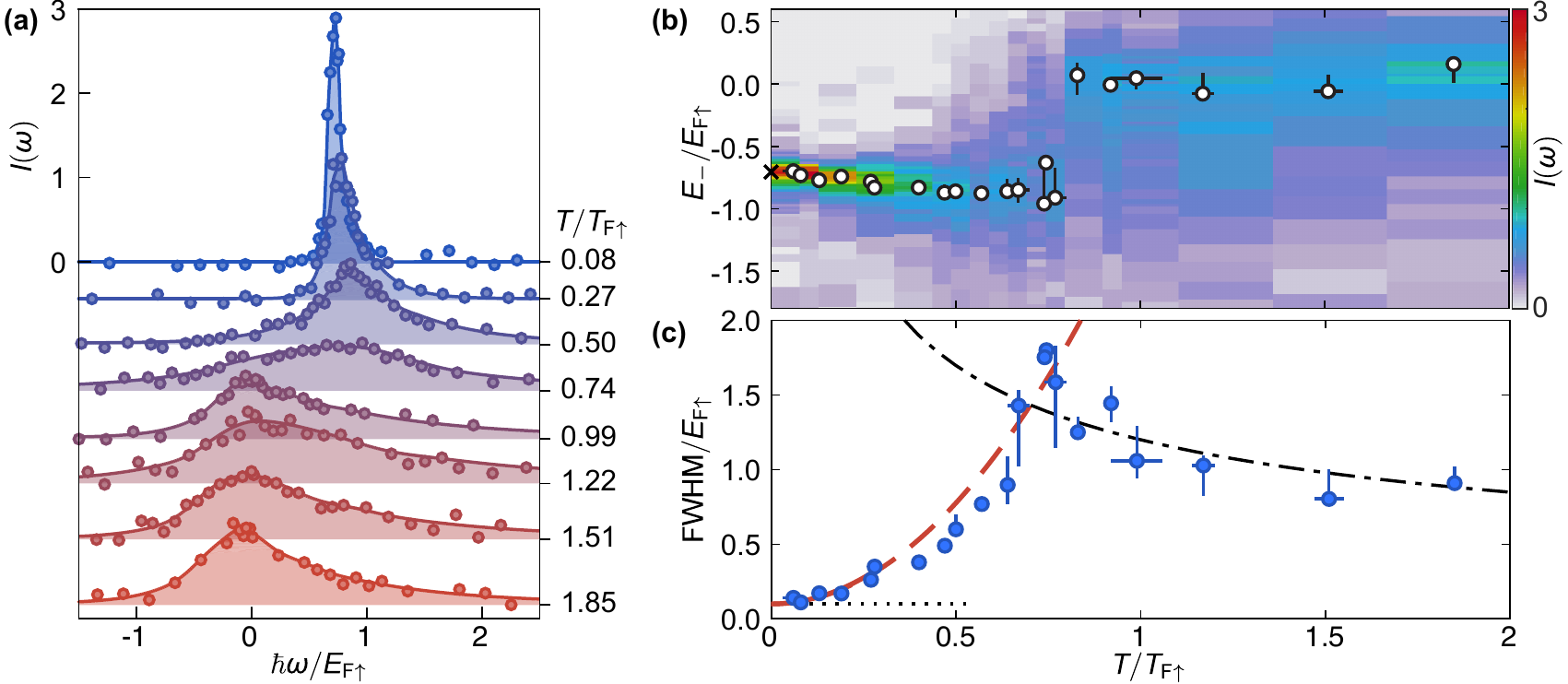}
\caption{(a) Thermal evolution of the minority rf spectra. The impurity concentration is $n_{\downarrow}/n_{\uparrow}=0.10\pm0.03$, the Rabi frequency $\Omega_{\mathrm{R}}=2\pi\cdot0.5\,\mathrm{kHz}$ and the pulse duration $T_{\mathrm{Pulse}}=1\,\mathrm{ms}$. (b) 2D plot of the minority spectra with maxima highlighted by white points. To reflect the energy of the initial many-body state, the spectra are shown with the inverse frequency $E_-/E_{\mathrm{F}\uparrow}$, where $E_-=-\hbar \omega$. The cross corresponds to the theoretical zero temperature result for the polaron energy, including a correction for final state interactions~\cite{Chevy2006a,Lobo2006d,Combescot2007b,Prokofev2008a,Massignan2014}. (c) FWHM of the rf spectra. (Dotted line) Fourier resolution limit; (dashed red line) single-polaron decay rate $\Gamma/E_{\mathrm{F}\uparrow}=2.71(T/T_{\mathrm{F}\uparrow})^2$~\cite{Bruun2008}, offset by the Fourier limit; (dash-dotted black line) FWHM of the rf spectrum in the high-temperature limit $\Gamma/E_{\mathrm{F}\uparrow}= 1.2\sqrt{T_{\mathrm{F}\uparrow}/T}$~\cite{Enss2011,Sun2015}, reflecting the scattering rate in the classical, unitary Boltzmann gas. For the errors in (b) and (c) see~\cite{Note1}.}
\label{fig:M1}
\end{figure*}

For the spectroscopic studies we employ rf \emph{ejection} spectroscopy, where the many-body state is first prepared and then probed by transferring a small fraction of one spin component into a weakly or noninteracting final state.
Radio-frequency ejection spectroscopy has been used to, e.g., measure interactions, correlations, pairing phenomena in Fermi gases~\cite{Zwierlein2014,Zwerger2016}, and more specifically, the binding energy of the attractive Fermi polaron at low temperatures~\cite{Schirotzek2009,Koschorreck2012}.
A prerequisite for our measurements is a spatially uniform box potential. This avoids the spectral broadening caused by an inhomogeneous density and impurity concentration~\cite{Mukherjee2017b,Note1}.
The three energetically lowest hyperfine states of $^{6}\mathrm{Li}$ (labeled $\ket{1}$, $\ket{2}$, $\ket{3}$) are utilized to create and probe the strongly interacting spin mixture. The minority (impurity) and majority components are prepared in $\ket{\downarrow}=\ket{1}$ and $\ket{\uparrow}=\ket{3}$ and transferred via the rf drive into the final state $\ket{\mathrm{f}}=\ket{2}$~\cite{Note1,Schunck2008}. All measurements have been performed at a magnetic field of $690\,\mathrm{G}$, where the interactions between minority and majority atoms are unitarity limited. Final state interactions are weakly repulsive with $k_{\mathrm{F}\uparrow}a_{\uparrow  \mathrm{f}} \lesssim 0.2$ ($a_{\uparrow  \mathrm{f}} = 62\,\mathrm{nm}$).
The impurity concentration (minority to majority density ratio $n_{\downarrow}/n_{\uparrow}$) is controllably varied between $10\%$ and $30\%$.

The rf response is linked to the probability that a hole of energy $E$ and momentum $\mathbf{p}$ is excited by ejecting a particle from the many-body state, as described by the occupied spectral function $\mathcal{A}_{-\downarrow}(\mathbf{p},E)$~\cite{Schirotzek2009,Massignan2014,Torma2016,Zwerger2016}.
Detecting a free particle of momentum $\mathbf{p}$ after rf transfer implies a momentum $\mathbf{p}$ and energy $E_\mathbf{p} = \mathbf{p}^2/2m-\mu_{\downarrow}-\hbar\omega$ of the leftover hole, where $\mu_{\downarrow}$ is the minority chemical potential and $\hbar \omega$ is the energy of the rf photon with respect to the noninteracting transition. The number of transferred minority atoms $N_f(\omega)$ is proportional to the momentum integral of the occupied spectral function $\mathcal{A}_{-\downarrow}(\mathbf{p},E_\mathbf{p})$.
Fermi liquids feature a spectral function that is sharply peaked around $\epsilon_{0} + \mathbf{p}^2/2m^{*} - \mu_{\downarrow}$, with the effective mass $m^{*}$ and dressed energy $\epsilon_{0}$ of the quasiparticles. The width of the peak is determined by the quasiparticle decay rate $\Gamma(p,T)$. For low temperatures and impurity concentrations only low-momentum states are populated and the peak position of the rf spectrum corresponds to the polaron binding energy~\cite{Schirotzek2009}.

Figure~\ref{fig:M1}(a) shows the evolution with temperature of the rf spectra. Here, we have defined the normalized transfer $I(\omega)=(N_{\mathrm{f}}(\omega)/N_{\downarrow}) (E_{\mathrm{F}\uparrow}/ \hbar \Omega_{\mathrm{R}}^2 T_{\mathrm{Pulse}})$, with the number of particles in the final (initial) state $N_{\mathrm{f}}$ ($N_{\downarrow}$), the pulse duration $T_{\mathrm{Pulse}}$ and the single particle Rabi frequency $\Omega_{\mathrm{R}}$. The term $\Omega_{\mathrm{R}}^2 T_{\mathrm{Pulse}}$ originates from the linear response to the rf pulse.
The factor $E_{\mathrm{F}\uparrow}/ \hbar$ in $I$ is owed to the scale invariance of the unitary Fermi gas, which implies that its spectral features, such as the peak position, amplitude, and width directly scale with the Fermi energy~\cite{Zwierlein2014,Zwerger2016}.
The normalized transfer only depends on the dimensionless parameters $T/T_{\mathrm{F}\uparrow}$, $n_{\downarrow}/n_{\uparrow}$, and $\hbar \omega/E_{\mathrm{F}\uparrow}$, apart from small corrections due to final state interactions and Fourier broadening that break the scale invariance of the system. The energy of the gas is measured by an isoenergetic release from the uniform to a harmonic trap. After thermalization, the in-trap size reveals the energy, from which we obtain the temperature via the equation of state~\cite{Note1}.

In the deeply degenerate limit $(T/T_{\mathrm{F \uparrow}}<0.1)$, we observe a sharply defined resonance [Fig. 1(a)] signaling the stable long-lived Fermi polaron~\cite{Schirotzek2009}. Its width, defined by the full width at half maximum (FWHM), is limited by the Fourier resolution.
From the position of the spectral peak at low temperature [Fig. 1(b)] and correction for weak final state interactions as in~\cite{Schirotzek2009}, we obtain a zero temperature polaron binding energy $A\equiv {\epsilon_{0}/E_{\mathrm{F}\uparrow}=-0.60\pm0.05}$, with a linear extrapolation of the peak positions below $T/T_{\mathrm{F}\uparrow}=0.3$.

With increasing temperature, the spectral peak initially shifts to higher frequencies and broadens significantly [Figs.~\ref{fig:M1}(b) and \ref{fig:M1}(c)]. A rise in the polaron binding energy with temperature is expected, given the increased scattering phase space of the majority spins, and is found theoretically~\cite{Tajima2018,Mulkerin2018}. However, note that the position of the maximum at finite temperature and impurity concentration is influenced by the density of states, the difference in the effective mass between initial and final state~\cite{Scazza2017}, and the thermal population of momentum states.
At a temperature near $T/T_{\mathrm{F}\uparrow} \approx 0.75$, a sharp jump in the position of the global maximum to $\omega \approx 0$ is observed [Figs.~\ref{fig:M1}(a) and \ref{fig:M1}(b)] \cite{Note2}. In this regime, the width of the spectra reaches its maximum [Fig.~\ref{fig:M1}(c)], on the order of the Fermi energy.
Beyond this temperature, the position of the maximum remains constant at $\omega \approx 0$, as expected theoretically~\cite{Enss2011,Sun2015}. It reflects a merging of attractive and repulsive branches, symmetric about zero on resonance~\cite{Ho2004a}, as the temperature exceeds their splitting. 

The spectral function of a Fermi liquid is a single Lorentzian peak with a width given by the decay rate of the quasiparticles~\cite{Nozieres1966}. The width of the rf spectra is dominated by this decay rate at low temperatures. We observe a quadratic scaling of the width at low temperatures, a hallmark of Fermi liquid theory, in agreement with a theoretical calculation [Fig.~\ref{fig:M1}(c)]~\cite{Bruun2008}. In the quantum critical regime around $T \approx T_{\mathrm{F}\uparrow}$, the lifetime of the polarons drops below the Fermi time ($h/E_{\mathrm{F}\uparrow}$), signaling a breakdown of quasiparticles~\cite{Nikolic2007,Enss2012,Frank2018}.
The decrease in width at temperatures beyond the Fermi temperature is expected for a classical Boltzmann gas with unitarity limited interactions. The thermal scattering rate in the dilute impurity limit is given by $\Gamma_{\mathrm{th}} = n_{\uparrow} \sigma_{\mathrm{th}} v_{\mathrm{th}} \sim 1/\sqrt{T} $, with the thermal velocity $v_{\mathrm{th}}\sim \sqrt{T}$, and the unitarity limited scattering cross section $\sigma_{\mathrm{th}} \sim {\lambda}^2  \sim 1/T$.


\begin{figure}
\centering
\includegraphics[width=8.6cm]{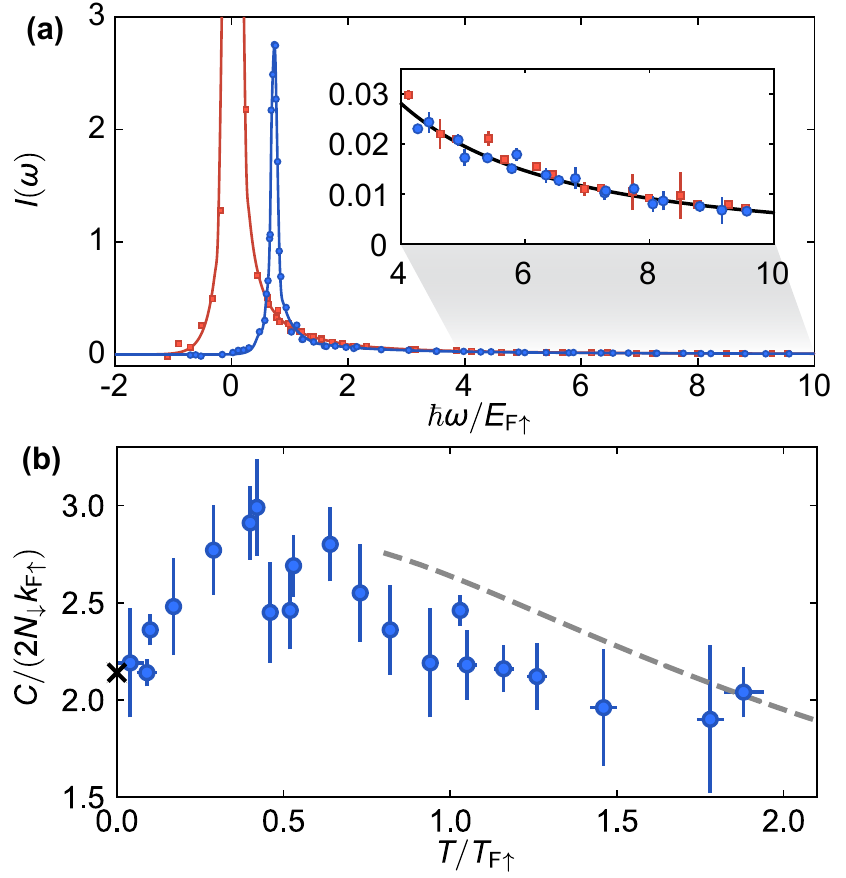}
\caption{Contact of the spin-imbalanced Fermi gas. (a) Typical rf spectra of the spin minority (blue circles) and majority (red squares). The impurity concentration is $10\%$. (Inset) High-frequency tails of the minority and majority spectra together with a fit of Eq.~\eqref{RFContact}. (b) Contact as a function of temperature, obtained by measuring the transferred fraction of atoms as a function of rf pulse duration for frequencies $\hbar \omega/E_{\mathrm{F} \uparrow} > 5.5$ and use of Eq.~\eqref{RFContact}. The gray dashed line shows the third-order viral expansion~\cite{Liu2010b} and the cross shows the result from the Chevy ansatz~\cite{Chevy2006a,Punk2009}.}
\label{fig:M2}
\end{figure}

Apart from energies and lifetimes, rf spectra also directly yield the strength of short-range correlations, quantified by contact $C$ [Fig~\ref{fig:M2}(a)]~\cite{Baym2007,Punk2007,Pieri2009,Schneider2009,Braaten2010,Stewart2010,Zwierlein2014}.
The contact is a central quantity in a set of universal relations, linking microscopic properties to thermodynamics, which apply to all many-body systems with contact interactions~\cite{Tan2008}. 
It governs the tail of the momentum distribution, short-range pair correlations and the change in energy with interaction strength~\cite{Braaten2010,Zwierlein2014,Zwerger2016}. 
As the contact is a measure of pair correlations, the tails of the rf spectrum of the minority and majority components are identical. 
For unitarity limited interactions the fraction of transferred atoms in the high-frequency limit is given by~\cite{Braaten2010}
\begin{equation}
I({\omega}) \underset{\omega \rightarrow \infty}{=} \frac{C}{2 N_{\downarrow}k_{\mathrm{F}\uparrow}} \frac{1}{2\sqrt{2}\pi\,(1+ \hbar \omega/E_{b})} \left(\frac{E_{\mathrm{F}\uparrow}}{\hbar \omega} \right)^{3/2},
\label{RFContact}
\end{equation}
where $E_{b}=\hbar^2/m a_{\uparrow  \mathrm{f}}^2 \approx h \cdot 433~\mathrm{kHz}$. The inset of Fig.~\ref {fig:M2}(a) shows the corresponding fit of the tails with Eq.~\eqref{RFContact}, leaving only the contact as a free parameter.

The temperature dependence of the contact displays a nonmonotonic behavior with a maximum located around $T\approx 0.4 \, T_{\mathrm{F}\uparrow}$ [Fig.~\ref {fig:M2}(b)]. The observed initial rise in temperature is partially expected from the increase in scattering phase space and has also been found theoretically in a spin-imbalanced few-body calculation of the contact \cite{Yan2013}.
In the high-temperature limit, the contact is proportional to the scattering cross section and vanishes as $1/T$.


\begin{figure}
\includegraphics[width=8.6cm]{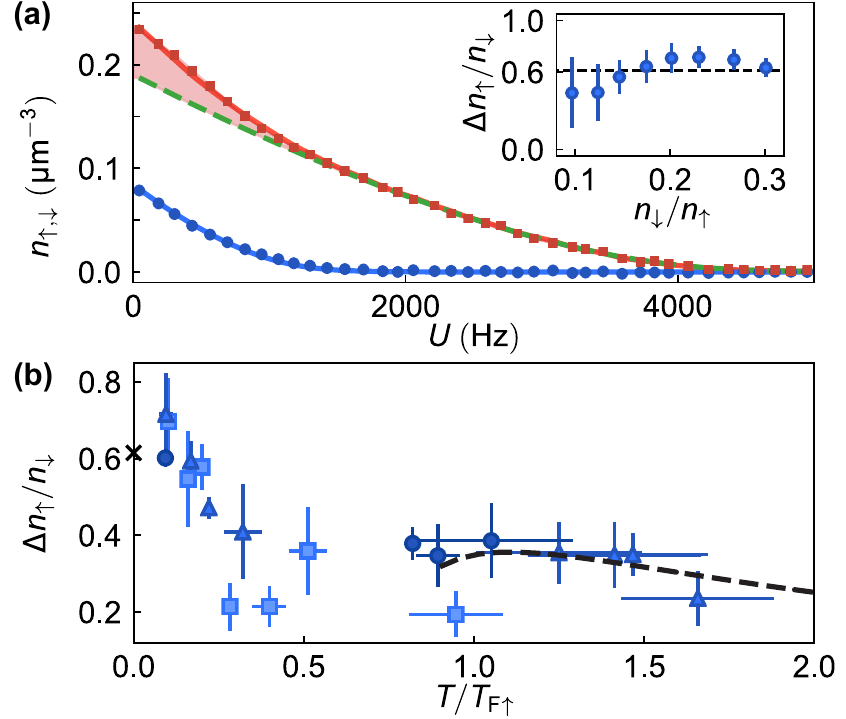}
\caption{Observation of the majority excess cloud. (a) Density profiles in a harmonically varying external potential $U$. Blue (red) data points indicate the profiles of the minority (majority) spin component. The normalized temperature of the gas is $T/T_{\mathrm{F}\uparrow}=0.07$ in the trap center ($U=0$). The green dashed line represents the equation of state of the ideal Fermi gas, the red (blue) solid line is the Fermi liquid ansatz [Eq.~\eqref{eq:density}] for the majority (minority) component. The red shaded area displays the excess majority density $\Delta n_{\uparrow}$. (Inset) Dependence of the excess majority to minority ratio on the impurity concentration. (b) Temperature dependence of the majority excess cloud. Data points show the excess majority density $\Delta n_{\uparrow}$ for an impurity concentration of $n_{\downarrow}/n_{\uparrow}=0.1$ (squares), $n_{\downarrow}/n_{\uparrow}=0.2$ (triangles), and $n_{\downarrow}/n_{\uparrow}=0.3$ (circles). The cross indicates the low-temperature prediction of the Fermi liquid ansatz $\Delta {n_{\uparrow}}/n_{\downarrow}=-A=0.615$~\cite{Prokofev2008a} and the dashed line shows the third-order virial expansion.}
\label{fig:M3}
\end{figure}

The contact quantifies short-range correlations. However, the polaron is an extended object with pair correlations extending out over distances even beyond the majority interparticle spacing~\cite{Trefzger2013}. We thus set out to probe the entire cloud of excess majority atoms surrounding the impurity spin of density $\Delta n_{\uparrow} = n_{\uparrow}(\mu_{\uparrow},\mu_{\downarrow},T) - n_{0}(\mu_{\uparrow},T)$ by \emph{in situ} density measurements [Fig.~\ref{fig:M3} (a)]. Here, $n_{\uparrow}(\mu_{\uparrow},\mu_{\downarrow},T)$ is the actual measured density of the interacting majority component and $n_{0}(\mu_{\uparrow},T)$ corresponds to the density of a noninteracting gas with the same temperature and majority chemical potential. For this measurement, we use a hybrid trapping potential that is harmonic along one direction and uniform along the other two axes~\cite{Mukherjee2017b}. This trapping geometry gives direct access to the density of each spin component as a function of the trapping potential $U$ [Fig.~\ref{fig:M3}(a)]. Under the local density approximation, the knowledge of $n_{\uparrow,\downarrow}(U)$ can be used to extract a variety of thermodynamic quantities~\cite{Nascimbene2010,Navon2010,Ku2012,Note1}.
The majority chemical potential and temperature are obtained from the low-fugacity wings of the gas. In the case of a partially spin polarized wing, we use the third-order virial expansion~\cite{Liu2010b}, whereas for a fully spin polarized wing, we use the ideal equation of state. For the lowest temperatures, the excess majority density per minority atom is $\Delta n_{\uparrow}/n_{\downarrow} = 0.63(5)$ [Fig.~\ref{fig:M3}(b)]. For increasing temperature, the excess density drops until it reaches the value predicted by the virial expansion for the density. $\Delta n_{\uparrow}/n_{\downarrow}$ displays no dependency on the minority concentration within our error up to $n_{\downarrow}/n_{\uparrow}=0.3$.

To elucidate the origin of the excess density from thermodynamics, we model the total pressure of the system as
\begin{equation}
P(\mu_{\uparrow},\mu_{\downarrow},T) = P_{0}(\mu_{\uparrow},T) + \left(\frac{m^*}{m^{\phantom{*}}}\right)^{\frac{3}{2}} P_{0}(\mu_{\downarrow}-A\mu_{\uparrow},T).
\end{equation}
Here, $P_{0}(\mu,T)$ is the pressure of the noninteracting Fermi gas. The ansatz describes the total pressure of the system as the sum of the partial pressure of the noninteracting majority component and the partial pressure of an ensemble of polarons with an effective chemical potential of $\mu_{\downarrow}-A\mu_{\uparrow}$ and an effective mass $m^*$~\cite{Nascimbene2010,Navon2010}. It contains weak interactions among the polarons that amount to a few percent of the total energy of the system \cite{Mora2010}. From this pressure ansatz the density can be calculated with the Gibbs-Duhem equation at constant temperature and scattering length $(\mathrm{d}P=n_{\uparrow} \mathrm{d}\mu_{\uparrow}+n_{\downarrow} \mathrm{d}\mu_{\downarrow})$,
\begin{eqnarray}
n_{\uparrow}(\mu_{\uparrow},\mu_{\downarrow},T)  &=& n_{0}(\mu_{\uparrow},T) - A n_{\downarrow}(\mu_{\uparrow},\mu_{\downarrow},T), \nonumber
\\
n_{\downarrow}(\mu_{\uparrow},\mu_{\downarrow},T)  &=& \left(m^*/m\right)^{\frac{3}{2}} n_{0}(\mu_{\downarrow}-A\mu_{\uparrow},T),
\label{eq:density}
\end{eqnarray}
where $n_{0}(\mu,T) \equiv \partial P_{0}/ \partial \mu$ is the density of the noninteracting gas. Each minority is accumulating on average $|A| = 0.6$ excess majority atoms over the noninteracting limit, in agreement with our measured value [Fig.~\ref{fig:M3}(b)].


\begin{figure}
\centering
\includegraphics[width=8.6cm]{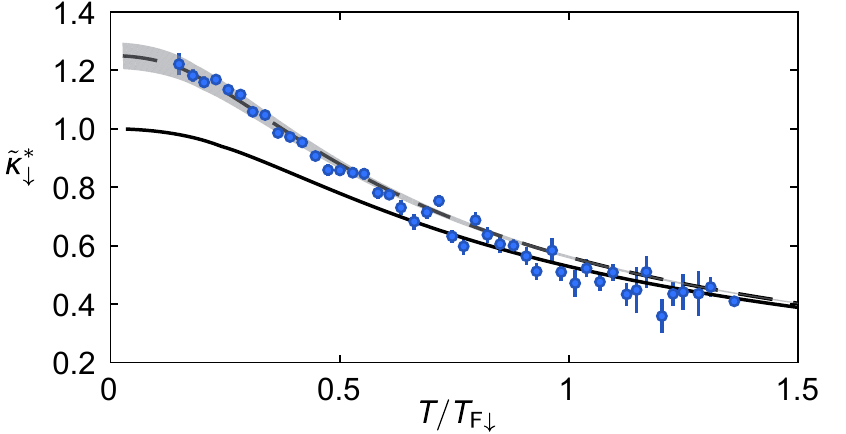}
\caption{Isothermal minority compressibility. The solid line is the Fermi liquid ansatz for $m^{*}/m=1$, while the dashed line corresponds to a fit with an effective mass of $m^{*}/m=1.25(5)$ assuming $A=-0.615$ ~\cite{Prokofev2008a}. The gray shaded area represents the standard deviation of the fit. For the entire range of temperatures displayed, the majority component is degenerate ($T/T_{\mathrm{F}\uparrow}<0.2$).}
\label{fig:M4}
\end{figure}

Since the Fermi liquid ansatz describes the thermodynamics accurately in the low-temperature regime $T/T_{\mathrm{F}\uparrow}<0.2$, we now focus on this temperature regime and utilize the ansatz to determine the effective mass of the polarons from a measurement of the minority compressibility.
In analogy to the total compressibility of the gas, the normalized isothermal minority compressibility is defined as $\tilde{\kappa}_{\downarrow} \equiv -\mathrm{d}E_{\mathrm{F\downarrow}}/\mathrm{d}U_{\mathrm{eff}}$~\cite{Ku2012}. Here, $U_{\mathrm{eff}}=(1-A)U$ is the effective potential of the minority component generated by the interaction with the majority component~\cite{Lobo2006d,Nascimbene2009}. Using Eq.~\eqref{eq:density} for the minority density, one finds
\begin{equation}
\tilde{\kappa}_{\downarrow} \left(T/T_{\mathrm{F}\downarrow}\right) = \frac{m^*}{m}\frac{\kappa_{0}\left(T,\,T_{\mathrm{F}\downarrow} \cdot m^{*}/m \right)}{\kappa_0 \left(0,\,T_{\mathrm{F}\downarrow}\cdot m^{*}/m \right)},
\label{eq:compressibility}
\end{equation}
where $\kappa_0 \left(T,\,T_{\mathrm{F}\downarrow} \right) \equiv n_{0}^{-2} \left(\partial n_{0}/ \partial \mu \right)_{T}$ is the compressibility of the noninteracting Fermi gas at fixed density. Figure~\ref{fig:M4} shows the measured isothermal compressibility of the minority component. A fit of Eq.~\eqref{eq:compressibility} fixing $A=-0.615$ ~\cite{Prokofev2008a} results in an effective mass of $m^{*}/m^{\phantom{*}}=1.25(5)$, which is in agreement with results obtained from diagrammatic Monte Carlo simulations~\cite{Prokofev2008a}, a variational ansatz~\cite{Combescot2007b}, and previous low-temperature experiments~\cite{Nascimbene2009,Nascimbene2010,Navon2010}. The saturation of the minority compressibility at low temperatures signals the formation of a degenerate Fermi sea of polarons.

In conclusion, we have studied the temperature dependence of a highly spin-imbalanced unitary Fermi gas with rf spectroscopy and in-trap density profiles. When the majority component is degenerate $(T/T_{\mathrm{F}\uparrow} \ll 1)$, long-lived quasiparticles emerge. In the spirit of Fermi liquid theory, these polarons behave like a weakly interacting Fermi gas forming a sharp Fermi sea for $T/T_{\mathrm{F}\downarrow} \ll 1$. The weakly interacting character of the quasiparticles is also reflected in the independence of the majority dressing cloud on the impurity concentration. In the opposing high-temperature regime, the gas is accurately described as a classical Boltzmann gas. At intermediate temperatures ($T \approx T_{\mathrm{F}\uparrow}$) the quasiparticle description breaks down. The spectral features of the attractive polarons dissolve, merging with excited branches such as dressed dimerons ~\cite{Prokofev2008a,Punk2009,Schmidt2011}, and repulsive polarons~\cite{Schmidt2011,Koschorreck2012,Kohstall2012a,Massignan2014,Scazza2017,Schmidt2018}.

\begin{acknowledgments}
We thank Richard Schmidt and Felix Werner for helpful discussions. This work was supported by the NSF, AFOSR, ONR, AFOSR MURI on Exotic Phases, and the David and Lucile Packard Foundation.
J.S. was supported by LabEX ENS-ICFP: ANR-10-LABX-0010/ANR-10-IDEX-0001-02 PSL*.
\end{acknowledgments}

%


\begin{thebibliography}{51}%
\makeatletter
\providecommand \@ifxundefined [1]{%
 \@ifx{#1\undefined}
}%
\providecommand \@ifnum [1]{%
 \ifnum #1\expandafter \@firstoftwo
 \else \expandafter \@secondoftwo
 \fi
}%
\providecommand \@ifx [1]{%
 \ifx #1\expandafter \@firstoftwo
 \else \expandafter \@secondoftwo
 \fi
}%
\providecommand \natexlab [1]{#1}%
\providecommand \enquote  [1]{``#1''}%
\providecommand \bibnamefont  [1]{#1}%
\providecommand \bibfnamefont [1]{#1}%
\providecommand \citenamefont [1]{#1}%
\providecommand \href@noop [0]{\@secondoftwo}%
\providecommand \href [0]{\begingroup \@sanitize@url \@href}%
\providecommand \@href[1]{\@@startlink{#1}\@@href}%
\providecommand \@@href[1]{\endgroup#1\@@endlink}%
\providecommand \@sanitize@url [0]{\catcode `\\12\catcode `\$12\catcode
  `\&12\catcode `\#12\catcode `\^12\catcode `\_12\catcode `\%12\relax}%
\providecommand \@@startlink[1]{}%
\providecommand \@@endlink[0]{}%
\providecommand \url  [0]{\begingroup\@sanitize@url \@url }%
\providecommand \@url [1]{\endgroup\@href {#1}{\urlprefix }}%
\providecommand \urlprefix  [0]{URL }%
\providecommand \Eprint [0]{\href }%
\providecommand \doibase [0]{http://dx.doi.org/}%
\providecommand \selectlanguage [0]{\@gobble}%
\providecommand \bibinfo  [0]{\@secondoftwo}%
\providecommand \bibfield  [0]{\@secondoftwo}%
\providecommand \translation [1]{[#1]}%
\providecommand \BibitemOpen [0]{}%
\providecommand \bibitemStop [0]{}%
\providecommand \bibitemNoStop [0]{.\EOS\space}%
\providecommand \EOS [0]{\spacefactor3000\relax}%
\providecommand \BibitemShut  [1]{\csname bibitem#1\endcsname}%
\let\auto@bib@innerbib\@empty
\bibitem [{\citenamefont {Nozi{\`{e}}res}\ and\ \citenamefont
  {Pines}(1966)}]{Nozieres1966}%
  \BibitemOpen
  \bibfield  {author} {\bibinfo {author} {\bibfnamefont {P.}~\bibnamefont
  {Nozi{\`{e}}res}}\ and\ \bibinfo {author} {\bibfnamefont {D.}~\bibnamefont
  {Pines}},\ }\href@noop {} {\emph {\bibinfo {title} {{The Theory of Quantum
  Liquids, Vol. I: Normal Fermi Liquids}}}},\ \bibinfo {edition} {1st}\ ed.\
  (\bibinfo  {publisher} {W.A. Benjamin},\ \bibinfo {address} {New York},\
  \bibinfo {year} {1966})\BibitemShut {NoStop}%
\bibitem [{\citenamefont {Zwierlein}(2006)}]{Zwierlein2006d}%
  \BibitemOpen
  \bibfield  {author} {\bibinfo {author} {\bibfnamefont {M.~W.}\ \bibnamefont
  {Zwierlein}},\ }\href {\doibase 10.1126/science.1122318} {\bibfield
  {journal} {\bibinfo  {journal} {Science}\ }\textbf {\bibinfo {volume}
  {311}},\ \bibinfo {pages} {492} (\bibinfo {year} {2006})}\BibitemShut
  {NoStop}%
\bibitem [{\citenamefont {Chevy}(2006)}]{Chevy2006a}%
  \BibitemOpen
  \bibfield  {author} {\bibinfo {author} {\bibfnamefont {F.}~\bibnamefont
  {Chevy}},\ }\href {\doibase 10.1103/PhysRevA.74.063628} {\bibfield  {journal}
  {\bibinfo  {journal} {Phys. Rev. A}\ }\textbf {\bibinfo {volume} {74}},\
  \bibinfo {pages} {063628} (\bibinfo {year} {2006})}\BibitemShut {NoStop}%
\bibitem [{\citenamefont {Lobo}\ \emph {et~al.}(2006)\citenamefont {Lobo},
  \citenamefont {Recati}, \citenamefont {Giorgini},\ and\ \citenamefont
  {Stringari}}]{Lobo2006d}%
  \BibitemOpen
  \bibfield  {author} {\bibinfo {author} {\bibfnamefont {C.}~\bibnamefont
  {Lobo}}, \bibinfo {author} {\bibfnamefont {A.}~\bibnamefont {Recati}},
  \bibinfo {author} {\bibfnamefont {S.}~\bibnamefont {Giorgini}}, \ and\
  \bibinfo {author} {\bibfnamefont {S.}~\bibnamefont {Stringari}},\ }\href
  {\doibase 10.1103/PhysRevLett.97.200403} {\bibfield  {journal} {\bibinfo
  {journal} {Phys. Rev. Lett.}\ }\textbf {\bibinfo {volume} {97}},\ \bibinfo
  {pages} {200403} (\bibinfo {year} {2006})}\BibitemShut {NoStop}%
\bibitem [{\citenamefont {Combescot}\ \emph {et~al.}(2007)\citenamefont
  {Combescot}, \citenamefont {Recati}, \citenamefont {Lobo},\ and\
  \citenamefont {Chevy}}]{Combescot2007b}%
  \BibitemOpen
  \bibfield  {author} {\bibinfo {author} {\bibfnamefont {R.}~\bibnamefont
  {Combescot}}, \bibinfo {author} {\bibfnamefont {A.}~\bibnamefont {Recati}},
  \bibinfo {author} {\bibfnamefont {C.}~\bibnamefont {Lobo}}, \ and\ \bibinfo
  {author} {\bibfnamefont {F.}~\bibnamefont {Chevy}},\ }\href {\doibase
  10.1103/PhysRevLett.98.180402} {\bibfield  {journal} {\bibinfo  {journal}
  {Phys. Rev. Lett.}\ }\textbf {\bibinfo {volume} {98}},\ \bibinfo {pages}
  {180402} (\bibinfo {year} {2007})}\BibitemShut {NoStop}%
\bibitem [{\citenamefont {Shin}\ \emph {et~al.}(2008)\citenamefont {Shin},
  \citenamefont {Schunck}, \citenamefont {Schirotzek},\ and\ \citenamefont
  {Ketterle}}]{Shin2008}%
  \BibitemOpen
  \bibfield  {author} {\bibinfo {author} {\bibfnamefont {Y.-I.}\ \bibnamefont
  {Shin}}, \bibinfo {author} {\bibfnamefont {C.~H.}\ \bibnamefont {Schunck}},
  \bibinfo {author} {\bibfnamefont {A.}~\bibnamefont {Schirotzek}}, \ and\
  \bibinfo {author} {\bibfnamefont {W.}~\bibnamefont {Ketterle}},\ }\href
  {\doibase 10.1038/nature06473} {\bibfield  {journal} {\bibinfo  {journal}
  {Nature (London)}\ }\textbf {\bibinfo {volume} {451}},\ \bibinfo {pages}
  {689} (\bibinfo {year} {2008})}\BibitemShut {NoStop}%
\bibitem [{\citenamefont {Bruun}\ \emph {et~al.}(2008)\citenamefont {Bruun},
  \citenamefont {Recati}, \citenamefont {Pethick}, \citenamefont {Smith},\ and\
  \citenamefont {Stringari}}]{Bruun2008}%
  \BibitemOpen
  \bibfield  {author} {\bibinfo {author} {\bibfnamefont {G.~M.}\ \bibnamefont
  {Bruun}}, \bibinfo {author} {\bibfnamefont {A.}~\bibnamefont {Recati}},
  \bibinfo {author} {\bibfnamefont {C.~J.}\ \bibnamefont {Pethick}}, \bibinfo
  {author} {\bibfnamefont {H.}~\bibnamefont {Smith}}, \ and\ \bibinfo {author}
  {\bibfnamefont {S.}~\bibnamefont {Stringari}},\ }\href {\doibase
  10.1103/PhysRevLett.100.240406} {\bibfield  {journal} {\bibinfo  {journal}
  {Phys. Rev. Lett.}\ }\textbf {\bibinfo {volume} {100}},\ \bibinfo {pages}
  {240406} (\bibinfo {year} {2008})}\BibitemShut {NoStop}%
\bibitem [{\citenamefont {Prokof'ev}\ and\ \citenamefont
  {Svistunov}(2008)}]{Prokofev2008a}%
  \BibitemOpen
  \bibfield  {author} {\bibinfo {author} {\bibfnamefont {N.}~\bibnamefont
  {Prokof'ev}}\ and\ \bibinfo {author} {\bibfnamefont {B.}~\bibnamefont
  {Svistunov}},\ }\href {\doibase 10.1103/PhysRevB.77.020408} {\bibfield
  {journal} {\bibinfo  {journal} {Phys. Rev. B}\ }\textbf {\bibinfo {volume}
  {77}},\ \bibinfo {pages} {020408} (\bibinfo {year} {2008})}\BibitemShut
  {NoStop}%
\bibitem [{\citenamefont {Veillette}\ \emph {et~al.}(2008)\citenamefont
  {Veillette}, \citenamefont {Moon}, \citenamefont {Lamacraft}, \citenamefont
  {Radzihovsky}, \citenamefont {Sachdev},\ and\ \citenamefont
  {Sheehy}}]{Veillette2008}%
  \BibitemOpen
  \bibfield  {author} {\bibinfo {author} {\bibfnamefont {M.}~\bibnamefont
  {Veillette}}, \bibinfo {author} {\bibfnamefont {E.~G.}\ \bibnamefont {Moon}},
  \bibinfo {author} {\bibfnamefont {A.}~\bibnamefont {Lamacraft}}, \bibinfo
  {author} {\bibfnamefont {L.}~\bibnamefont {Radzihovsky}}, \bibinfo {author}
  {\bibfnamefont {S.}~\bibnamefont {Sachdev}}, \ and\ \bibinfo {author}
  {\bibfnamefont {D.~E.}\ \bibnamefont {Sheehy}},\ }\href {\doibase
  10.1103/PhysRevA.78.033614} {\bibfield  {journal} {\bibinfo  {journal} {Phys.
  Rev. A}\ }\textbf {\bibinfo {volume} {78}},\ \bibinfo {pages} {033614}
  (\bibinfo {year} {2008})}\BibitemShut {NoStop}%
\bibitem [{\citenamefont {Schirotzek}\ \emph {et~al.}(2009)\citenamefont
  {Schirotzek}, \citenamefont {Wu}, \citenamefont {Sommer},\ and\ \citenamefont
  {Zwierlein}}]{Schirotzek2009}%
  \BibitemOpen
  \bibfield  {author} {\bibinfo {author} {\bibfnamefont {A.}~\bibnamefont
  {Schirotzek}}, \bibinfo {author} {\bibfnamefont {C.-H.}\ \bibnamefont {Wu}},
  \bibinfo {author} {\bibfnamefont {A.}~\bibnamefont {Sommer}}, \ and\ \bibinfo
  {author} {\bibfnamefont {M.~W.}\ \bibnamefont {Zwierlein}},\ }\href {\doibase
  10.1103/PhysRevLett.102.230402} {\bibfield  {journal} {\bibinfo  {journal}
  {Phys. Rev. Lett.}\ }\textbf {\bibinfo {volume} {102}},\ \bibinfo {pages}
  {230402} (\bibinfo {year} {2009})}\BibitemShut {NoStop}%
\bibitem [{\citenamefont {Nascimb{\`{e}}ne}\ \emph {et~al.}(2009)\citenamefont
  {Nascimb{\`{e}}ne}, \citenamefont {Navon}, \citenamefont {Jiang},
  \citenamefont {Tarruell}, \citenamefont {Teichmann}, \citenamefont
  {McKeever}, \citenamefont {Chevy},\ and\ \citenamefont
  {Salomon}}]{Nascimbene2009}%
  \BibitemOpen
  \bibfield  {author} {\bibinfo {author} {\bibfnamefont {S.}~\bibnamefont
  {Nascimb{\`{e}}ne}}, \bibinfo {author} {\bibfnamefont {N.}~\bibnamefont
  {Navon}}, \bibinfo {author} {\bibfnamefont {K.~J.}\ \bibnamefont {Jiang}},
  \bibinfo {author} {\bibfnamefont {L.}~\bibnamefont {Tarruell}}, \bibinfo
  {author} {\bibfnamefont {M.}~\bibnamefont {Teichmann}}, \bibinfo {author}
  {\bibfnamefont {J.}~\bibnamefont {McKeever}}, \bibinfo {author}
  {\bibfnamefont {F.}~\bibnamefont {Chevy}}, \ and\ \bibinfo {author}
  {\bibfnamefont {C.}~\bibnamefont {Salomon}},\ }\href {\doibase
  10.1103/PhysRevLett.103.170402} {\bibfield  {journal} {\bibinfo  {journal}
  {Phys. Rev. Lett.}\ }\textbf {\bibinfo {volume} {103}},\ \bibinfo {pages}
  {170402} (\bibinfo {year} {2009})}\BibitemShut {NoStop}%
\bibitem [{\citenamefont {Nascimb{\`{e}}ne}\ \emph {et~al.}(2010)\citenamefont
  {Nascimb{\`{e}}ne}, \citenamefont {Navon}, \citenamefont {Jiang},
  \citenamefont {Chevy},\ and\ \citenamefont {Salomon}}]{Nascimbene2010}%
  \BibitemOpen
  \bibfield  {author} {\bibinfo {author} {\bibfnamefont {S.}~\bibnamefont
  {Nascimb{\`{e}}ne}}, \bibinfo {author} {\bibfnamefont {N.}~\bibnamefont
  {Navon}}, \bibinfo {author} {\bibfnamefont {K.~J.}\ \bibnamefont {Jiang}},
  \bibinfo {author} {\bibfnamefont {F.}~\bibnamefont {Chevy}}, \ and\ \bibinfo
  {author} {\bibfnamefont {C.}~\bibnamefont {Salomon}},\ }\href {\doibase
  10.1038/nature08814} {\bibfield  {journal} {\bibinfo  {journal} {Nature
  (London)}\ }\textbf {\bibinfo {volume} {463}},\ \bibinfo {pages} {1057}
  (\bibinfo {year} {2010})}\BibitemShut {NoStop}%
\bibitem [{\citenamefont {Navon}\ \emph {et~al.}(2010)\citenamefont {Navon},
  \citenamefont {Nascimbene}, \citenamefont {Chevy},\ and\ \citenamefont
  {Salomon}}]{Navon2010}%
  \BibitemOpen
  \bibfield  {author} {\bibinfo {author} {\bibfnamefont {N.}~\bibnamefont
  {Navon}}, \bibinfo {author} {\bibfnamefont {S.}~\bibnamefont {Nascimbene}},
  \bibinfo {author} {\bibfnamefont {F.}~\bibnamefont {Chevy}}, \ and\ \bibinfo
  {author} {\bibfnamefont {C.}~\bibnamefont {Salomon}},\ }\href {\doibase
  10.1126/science.1187582} {\bibfield  {journal} {\bibinfo  {journal}
  {Science}\ }\textbf {\bibinfo {volume} {328}},\ \bibinfo {pages} {729}
  (\bibinfo {year} {2010})}\BibitemShut {NoStop}%
\bibitem [{\citenamefont {Sommer}\ \emph {et~al.}(2011)\citenamefont {Sommer},
  \citenamefont {Ku},\ and\ \citenamefont {Zwierlein}}]{Sommer2011b}%
  \BibitemOpen
  \bibfield  {author} {\bibinfo {author} {\bibfnamefont {A.}~\bibnamefont
  {Sommer}}, \bibinfo {author} {\bibfnamefont {M.}~\bibnamefont {Ku}}, \ and\
  \bibinfo {author} {\bibfnamefont {M.~W.}\ \bibnamefont {Zwierlein}},\ }\href
  {\doibase 10.1088/1367-2630/13/5/055009} {\bibfield  {journal} {\bibinfo
  {journal} {New J. Phys.}\ }\textbf {\bibinfo {volume} {13}},\ \bibinfo
  {pages} {055009} (\bibinfo {year} {2011})}\BibitemShut {NoStop}%
\bibitem [{\citenamefont {Nikoli{\'{c}}}\ and\ \citenamefont
  {Sachdev}(2007)}]{Nikolic2007}%
  \BibitemOpen
  \bibfield  {author} {\bibinfo {author} {\bibfnamefont {P.}~\bibnamefont
  {Nikoli{\'{c}}}}\ and\ \bibinfo {author} {\bibfnamefont {S.}~\bibnamefont
  {Sachdev}},\ }\href {\doibase 10.1103/PhysRevA.75.033608} {\bibfield
  {journal} {\bibinfo  {journal} {Phys. Rev. A}\ }\textbf {\bibinfo {volume}
  {75}},\ \bibinfo {pages} {033608} (\bibinfo {year} {2007})}\BibitemShut
  {NoStop}%
\bibitem [{\citenamefont {Enss}(2012)}]{Enss2012}%
  \BibitemOpen
  \bibfield  {author} {\bibinfo {author} {\bibfnamefont {T.}~\bibnamefont
  {Enss}},\ }\href {\doibase 10.1103/PhysRevA.86.013616} {\bibfield  {journal}
  {\bibinfo  {journal} {Phys. Rev. A}\ }\textbf {\bibinfo {volume} {86}},\
  \bibinfo {pages} {013616} (\bibinfo {year} {2012})}\BibitemShut {NoStop}%
\bibitem [{\citenamefont {Frank}\ \emph {et~al.}(2018)\citenamefont {Frank},
  \citenamefont {Lang},\ and\ \citenamefont {Zwerger}}]{Frank2018}%
  \BibitemOpen
  \bibfield  {author} {\bibinfo {author} {\bibfnamefont {B.}~\bibnamefont
  {Frank}}, \bibinfo {author} {\bibfnamefont {J.}~\bibnamefont {Lang}}, \ and\
  \bibinfo {author} {\bibfnamefont {W.}~\bibnamefont {Zwerger}},\ }\href
  {http://arxiv.org/abs/1804.03035} {\bibfield  {journal} {\bibinfo  {journal}
  {J. Exp. Theor. Phys.}\ }\textbf {\bibinfo {volume} {127}},\ \bibinfo {pages}
  {812} (\bibinfo {year} {2018})}\BibitemShut {NoStop}%
\bibitem [{\citenamefont {Kohstall}\ \emph {et~al.}(2012)\citenamefont
  {Kohstall}, \citenamefont {Zaccanti}, \citenamefont {Jag}, \citenamefont
  {Trenkwalder}, \citenamefont {Massignan}, \citenamefont {Bruun},
  \citenamefont {Schreck},\ and\ \citenamefont {Grimm}}]{Kohstall2012a}%
  \BibitemOpen
  \bibfield  {author} {\bibinfo {author} {\bibfnamefont {C.}~\bibnamefont
  {Kohstall}}, \bibinfo {author} {\bibfnamefont {M.}~\bibnamefont {Zaccanti}},
  \bibinfo {author} {\bibfnamefont {M.}~\bibnamefont {Jag}}, \bibinfo {author}
  {\bibfnamefont {A.}~\bibnamefont {Trenkwalder}}, \bibinfo {author}
  {\bibfnamefont {P.}~\bibnamefont {Massignan}}, \bibinfo {author}
  {\bibfnamefont {G.~M.}\ \bibnamefont {Bruun}}, \bibinfo {author}
  {\bibfnamefont {F.}~\bibnamefont {Schreck}}, \ and\ \bibinfo {author}
  {\bibfnamefont {R.}~\bibnamefont {Grimm}},\ }\href {\doibase
  10.1038/nature11065} {\bibfield  {journal} {\bibinfo  {journal} {Nature
  (London)}\ }\textbf {\bibinfo {volume} {485}},\ \bibinfo {pages} {615}
  (\bibinfo {year} {2012})}\BibitemShut {NoStop}%
\bibitem [{\citenamefont {Koschorreck}\ \emph {et~al.}(2012)\citenamefont
  {Koschorreck}, \citenamefont {Pertot}, \citenamefont {Vogt}, \citenamefont
  {Fr{\"{o}}hlich}, \citenamefont {Feld},\ and\ \citenamefont
  {K{\"{o}}hl}}]{Koschorreck2012}%
  \BibitemOpen
  \bibfield  {author} {\bibinfo {author} {\bibfnamefont {M.}~\bibnamefont
  {Koschorreck}}, \bibinfo {author} {\bibfnamefont {D.}~\bibnamefont {Pertot}},
  \bibinfo {author} {\bibfnamefont {E.}~\bibnamefont {Vogt}}, \bibinfo {author}
  {\bibfnamefont {B.}~\bibnamefont {Fr{\"{o}}hlich}}, \bibinfo {author}
  {\bibfnamefont {M.}~\bibnamefont {Feld}}, \ and\ \bibinfo {author}
  {\bibfnamefont {M.}~\bibnamefont {K{\"{o}}hl}},\ }\href {\doibase
  10.1038/nature11151} {\bibfield  {journal} {\bibinfo  {journal} {Nature
  (London)}\ }\textbf {\bibinfo {volume} {485}},\ \bibinfo {pages} {619}
  (\bibinfo {year} {2012})}\BibitemShut {NoStop}%
\bibitem [{\citenamefont {Scazza}\ \emph {et~al.}(2017)\citenamefont {Scazza},
  \citenamefont {Valtolina}, \citenamefont {Massignan}, \citenamefont {Recati},
  \citenamefont {Amico}, \citenamefont {Burchianti}, \citenamefont {Fort},
  \citenamefont {Inguscio}, \citenamefont {Zaccanti},\ and\ \citenamefont
  {Roati}}]{Scazza2017}%
  \BibitemOpen
  \bibfield  {author} {\bibinfo {author} {\bibfnamefont {F.}~\bibnamefont
  {Scazza}}, \bibinfo {author} {\bibfnamefont {G.}~\bibnamefont {Valtolina}},
  \bibinfo {author} {\bibfnamefont {P.}~\bibnamefont {Massignan}}, \bibinfo
  {author} {\bibfnamefont {A.}~\bibnamefont {Recati}}, \bibinfo {author}
  {\bibfnamefont {A.}~\bibnamefont {Amico}}, \bibinfo {author} {\bibfnamefont
  {A.}~\bibnamefont {Burchianti}}, \bibinfo {author} {\bibfnamefont
  {C.}~\bibnamefont {Fort}}, \bibinfo {author} {\bibfnamefont {M.}~\bibnamefont
  {Inguscio}}, \bibinfo {author} {\bibfnamefont {M.}~\bibnamefont {Zaccanti}},
  \ and\ \bibinfo {author} {\bibfnamefont {G.}~\bibnamefont {Roati}},\ }\href
  {\doibase 10.1103/PhysRevLett.118.083602} {\bibfield  {journal} {\bibinfo
  {journal} {Phys. Rev. Lett.}\ }\textbf {\bibinfo {volume} {118}},\ \bibinfo
  {pages} {083602} (\bibinfo {year} {2017})}\BibitemShut {NoStop}%
\bibitem [{\citenamefont {Mukherjee}\ \emph {et~al.}(2017)\citenamefont
  {Mukherjee}, \citenamefont {Yan}, \citenamefont {Patel}, \citenamefont
  {Hadzibabic}, \citenamefont {Yefsah}, \citenamefont {Struck},\ and\
  \citenamefont {Zwierlein}}]{Mukherjee2017b}%
  \BibitemOpen
  \bibfield  {author} {\bibinfo {author} {\bibfnamefont {B.}~\bibnamefont
  {Mukherjee}}, \bibinfo {author} {\bibfnamefont {Z.}~\bibnamefont {Yan}},
  \bibinfo {author} {\bibfnamefont {P.~B.}\ \bibnamefont {Patel}}, \bibinfo
  {author} {\bibfnamefont {Z.}~\bibnamefont {Hadzibabic}}, \bibinfo {author}
  {\bibfnamefont {T.}~\bibnamefont {Yefsah}}, \bibinfo {author} {\bibfnamefont
  {J.}~\bibnamefont {Struck}}, \ and\ \bibinfo {author} {\bibfnamefont {M.~W.}\
  \bibnamefont {Zwierlein}},\ }\href {\doibase 10.1103/PhysRevLett.118.123401}
  {\bibfield  {journal} {\bibinfo  {journal} {Phys. Rev. Lett.}\ }\textbf
  {\bibinfo {volume} {118}},\ \bibinfo {pages} {123401} (\bibinfo {year}
  {2017})}\BibitemShut {NoStop}%
\bibitem [{\citenamefont {Hueck}\ \emph {et~al.}(2018)\citenamefont {Hueck},
  \citenamefont {Luick}, \citenamefont {Sobirey}, \citenamefont {Siegl},
  \citenamefont {Lompe},\ and\ \citenamefont {Moritz}}]{Hueck2018}%
  \BibitemOpen
  \bibfield  {author} {\bibinfo {author} {\bibfnamefont {K.}~\bibnamefont
  {Hueck}}, \bibinfo {author} {\bibfnamefont {N.}~\bibnamefont {Luick}},
  \bibinfo {author} {\bibfnamefont {L.}~\bibnamefont {Sobirey}}, \bibinfo
  {author} {\bibfnamefont {J.}~\bibnamefont {Siegl}}, \bibinfo {author}
  {\bibfnamefont {T.}~\bibnamefont {Lompe}}, \ and\ \bibinfo {author}
  {\bibfnamefont {H.}~\bibnamefont {Moritz}},\ }\href {\doibase
  10.1103/PhysRevLett.120.060402} {\bibfield  {journal} {\bibinfo  {journal}
  {Phys. Rev. Lett.}\ }\textbf {\bibinfo {volume} {120}},\ \bibinfo {pages}
  {060402} (\bibinfo {year} {2018})}\BibitemShut {NoStop}%
\bibitem [{\citenamefont {Tan}(2008)}]{Tan2008}%
  \BibitemOpen
  \bibfield  {author} {\bibinfo {author} {\bibfnamefont {S.}~\bibnamefont
  {Tan}},\ }\href {\doibase 10.1016/j.aop.2008.03.005} {\bibfield  {journal}
  {\bibinfo  {journal} {Ann. Phys. (Amsterdam)}\ }\textbf {\bibinfo {volume}
  {323}},\ \bibinfo {pages} {2971} (\bibinfo {year} {2008})}\BibitemShut
  {NoStop}%
\bibitem [{\citenamefont {Baym}\ \emph {et~al.}(2007)\citenamefont {Baym},
  \citenamefont {Pethick}, \citenamefont {Yu},\ and\ \citenamefont
  {Zwierlein}}]{Baym2007}%
  \BibitemOpen
  \bibfield  {author} {\bibinfo {author} {\bibfnamefont {G.}~\bibnamefont
  {Baym}}, \bibinfo {author} {\bibfnamefont {C.~J.}\ \bibnamefont {Pethick}},
  \bibinfo {author} {\bibfnamefont {Z.}~\bibnamefont {Yu}}, \ and\ \bibinfo
  {author} {\bibfnamefont {M.~W.}\ \bibnamefont {Zwierlein}},\ }\href {\doibase
  10.1103/PhysRevLett.99.190407} {\bibfield  {journal} {\bibinfo  {journal}
  {Phys. Rev. Lett.}\ }\textbf {\bibinfo {volume} {99}},\ \bibinfo {pages}
  {190407} (\bibinfo {year} {2007})}\BibitemShut {NoStop}%
\bibitem [{\citenamefont {Punk}\ and\ \citenamefont
  {Zwerger}(2007)}]{Punk2007}%
  \BibitemOpen
  \bibfield  {author} {\bibinfo {author} {\bibfnamefont {M.}~\bibnamefont
  {Punk}}\ and\ \bibinfo {author} {\bibfnamefont {W.}~\bibnamefont {Zwerger}},\
  }\href {\doibase 10.1103/PhysRevLett.99.170404} {\bibfield  {journal}
  {\bibinfo  {journal} {Phys. Rev. Lett.}\ }\textbf {\bibinfo {volume} {99}},\
  \bibinfo {pages} {170404} (\bibinfo {year} {2007})}\BibitemShut {NoStop}%
\bibitem [{\citenamefont {Schneider}\ \emph {et~al.}()\citenamefont
  {Schneider}, \citenamefont {Shenoy},\ and\ \citenamefont
  {Randeria}}]{Schneider2009}%
  \BibitemOpen
  \bibfield  {author} {\bibinfo {author} {\bibfnamefont {W.}~\bibnamefont
  {Schneider}}, \bibinfo {author} {\bibfnamefont {V.~B.}\ \bibnamefont
  {Shenoy}}, \ and\ \bibinfo {author} {\bibfnamefont {M.}~\bibnamefont
  {Randeria}},\ }\href {http://arxiv.org/abs/0903.3006} {\bibinfo  {journal}
  {arXiv:0903.3006}}\BibitemShut {NoStop}%
\bibitem [{\citenamefont {Braaten}\ \emph {et~al.}(2010)\citenamefont
  {Braaten}, \citenamefont {Kang},\ and\ \citenamefont
  {Platter}}]{Braaten2010}%
  \BibitemOpen
\bibfield  {journal} {  }\bibfield  {author} {\bibinfo {author} {\bibfnamefont
  {E.}~\bibnamefont {Braaten}}, \bibinfo {author} {\bibfnamefont
  {D.}~\bibnamefont {Kang}}, \ and\ \bibinfo {author} {\bibfnamefont
  {L.}~\bibnamefont {Platter}},\ }\href {\doibase
  10.1103/PhysRevLett.104.223004} {\bibfield  {journal} {\bibinfo  {journal}
  {Phys. Rev. Lett.}\ }\textbf {\bibinfo {volume} {104}},\ \bibinfo {pages}
  {223004} (\bibinfo {year} {2010})}\BibitemShut {NoStop}%
\bibitem [{\citenamefont {Shin}(2008)}]{Shin2008a}%
  \BibitemOpen
  \bibfield  {author} {\bibinfo {author} {\bibfnamefont {Y.-I.}\ \bibnamefont
  {Shin}},\ }\href {\doibase 10.1103/PhysRevA.77.041603} {\bibfield  {journal}
  {\bibinfo  {journal} {Phys. Rev. A}\ }\textbf {\bibinfo {volume} {77}},\
  \bibinfo {pages} {041603} (\bibinfo {year} {2008})}\BibitemShut {NoStop}%
\bibitem [{\citenamefont {Horikoshi}\ \emph {et~al.}(2010)\citenamefont
  {Horikoshi}, \citenamefont {Nakajima}, \citenamefont {Ueda},\ and\
  \citenamefont {Mukaiyama}}]{Horikoshi2010}%
  \BibitemOpen
  \bibfield  {author} {\bibinfo {author} {\bibfnamefont {M.}~\bibnamefont
  {Horikoshi}}, \bibinfo {author} {\bibfnamefont {S.}~\bibnamefont {Nakajima}},
  \bibinfo {author} {\bibfnamefont {M.}~\bibnamefont {Ueda}}, \ and\ \bibinfo
  {author} {\bibfnamefont {T.}~\bibnamefont {Mukaiyama}},\ }\href {\doibase
  10.1126/science.1183012} {\bibfield  {journal} {\bibinfo  {journal}
  {Science}\ }\textbf {\bibinfo {volume} {327}},\ \bibinfo {pages} {442}
  (\bibinfo {year} {2010})}\BibitemShut {NoStop}%
\bibitem [{\citenamefont {Ku}\ \emph {et~al.}(2012)\citenamefont {Ku},
  \citenamefont {Sommer}, \citenamefont {Cheuk},\ and\ \citenamefont
  {Zwierlein}}]{Ku2012}%
  \BibitemOpen
  \bibfield  {author} {\bibinfo {author} {\bibfnamefont {M.~J.~H.}\
  \bibnamefont {Ku}}, \bibinfo {author} {\bibfnamefont {A.~T.}\ \bibnamefont
  {Sommer}}, \bibinfo {author} {\bibfnamefont {L.~W.}\ \bibnamefont {Cheuk}}, \
  and\ \bibinfo {author} {\bibfnamefont {M.~W.}\ \bibnamefont {Zwierlein}},\
  }\href {\doibase 10.1126/science.1214987} {\bibfield  {journal} {\bibinfo
  {journal} {Science}\ }\textbf {\bibinfo {volume} {335}},\ \bibinfo {pages}
  {563} (\bibinfo {year} {2012})}\BibitemShut {NoStop}%
\bibitem [{\citenamefont {Zwierlein}(2014)}]{Zwierlein2014}%
  \BibitemOpen
  \bibfield  {author} {\bibinfo {author} {\bibfnamefont {M.~W.}\ \bibnamefont
  {Zwierlein}},\ }in\ \href {\doibase
  10.1093/acprof:oso/9780198719267.003.0007} {\emph {\bibinfo {booktitle}
  {Novel Superfluids}}},\ \bibinfo {editor} {edited by\ \bibinfo {editor}
  {\bibfnamefont {K.-H.}\ \bibnamefont {Bennemann}}\ and\ \bibinfo {editor}
  {\bibfnamefont {J.~B.}\ \bibnamefont {Ketterson}}}\ (\bibinfo  {publisher}
  {Oxford University Press},\ \bibinfo {address} {New York},\ \bibinfo {year}
  {2014})\ \bibinfo {edition} {1st}\ ed.,\ Chap.~\bibinfo {chapter} {18}, pp.\
  \bibinfo {pages} {269--422}\BibitemShut {NoStop}%
\bibitem [{\citenamefont {Zwerger}(2016)}]{Zwerger2016}%
  \BibitemOpen
  \bibfield  {author} {\bibinfo {author} {\bibfnamefont {W.}~\bibnamefont
  {Zwerger}},\ }in\ \href {\doibase 10.3254/978-1-61499-694-1-63} {\emph
  {\bibinfo {booktitle} {Quantum Matter at Ultralow Temperatures, Proceedings
  of the International School of Physics "Enrico Fermi", Course CXCI}}},\
  \bibinfo {editor} {edited by\ \bibinfo {editor} {\bibfnamefont
  {M.}~\bibnamefont {Inguscio}}, \bibinfo {editor} {\bibfnamefont
  {W.}~\bibnamefont {Ketterle}}, \bibinfo {editor} {\bibfnamefont
  {S.}~\bibnamefont {Stringari}}, \ and\ \bibinfo {editor} {\bibfnamefont
  {G.}~\bibnamefont {Roati}}}\ (\bibinfo  {publisher} {IOS Press},\ \bibinfo
  {address} {Amsterdam},\ \bibinfo {year} {2016})\ Chap.~\bibinfo {chapter}
  {2}, pp.\ \bibinfo {pages} {63--142}\BibitemShut {NoStop}%
\bibitem [{Note1()}]{Note1}%
  \BibitemOpen
  \bibinfo {note} {See attached supplemental material for more information
  regarding the state preparation, homogeneity of the gas, thermometry, and rf
  spectroscopy.}\BibitemShut {Stop}%
\bibitem [{\citenamefont {Schunck}\ \emph {et~al.}(2008)\citenamefont
  {Schunck}, \citenamefont {Shin}, \citenamefont {Schirotzek},\ and\
  \citenamefont {Ketterle}}]{Schunck2008}%
  \BibitemOpen
  \bibfield  {author} {\bibinfo {author} {\bibfnamefont {C.~H.}\ \bibnamefont
  {Schunck}}, \bibinfo {author} {\bibfnamefont {Y.-I.}\ \bibnamefont {Shin}},
  \bibinfo {author} {\bibfnamefont {A.}~\bibnamefont {Schirotzek}}, \ and\
  \bibinfo {author} {\bibfnamefont {W.}~\bibnamefont {Ketterle}},\ }\href
  {\doibase 10.1038/nature07176} {\bibfield  {journal} {\bibinfo  {journal}
  {Nature (London)}\ }\textbf {\bibinfo {volume} {454}},\ \bibinfo {pages}
  {739} (\bibinfo {year} {2008})}\BibitemShut {NoStop}%
\bibitem [{\citenamefont {Massignan}\ \emph {et~al.}(2014)\citenamefont
  {Massignan}, \citenamefont {Zaccanti},\ and\ \citenamefont
  {Bruun}}]{Massignan2014}%
  \BibitemOpen
  \bibfield  {author} {\bibinfo {author} {\bibfnamefont {P.}~\bibnamefont
  {Massignan}}, \bibinfo {author} {\bibfnamefont {M.}~\bibnamefont {Zaccanti}},
  \ and\ \bibinfo {author} {\bibfnamefont {G.~M.}\ \bibnamefont {Bruun}},\
  }\href {\doibase 10.1088/0034-4885/77/3/034401} {\bibfield  {journal}
  {\bibinfo  {journal} {Rep. Prog. Phys.}\ }\textbf {\bibinfo {volume} {77}},\
  \bibinfo {pages} {034401} (\bibinfo {year} {2014})}\BibitemShut {NoStop}%
\bibitem [{\citenamefont {T{\"{o}}rm{\"{a}}}(2016)}]{Torma2016}%
  \BibitemOpen
  \bibfield  {author} {\bibinfo {author} {\bibfnamefont {P.}~\bibnamefont
  {T{\"{o}}rm{\"{a}}}},\ }\href {\doibase 10.1088/0031-8949/91/4/043006}
  {\bibfield  {journal} {\bibinfo  {journal} {Phys. Scr.}\ }\textbf {\bibinfo
  {volume} {91}},\ \bibinfo {pages} {043006} (\bibinfo {year}
  {2016})}\BibitemShut {NoStop}%
\bibitem [{\citenamefont {Tajima}\ and\ \citenamefont
  {Uchino}(2018)}]{Tajima2018}%
  \BibitemOpen
  \bibfield  {author} {\bibinfo {author} {\bibfnamefont {H.}~\bibnamefont
  {Tajima}}\ and\ \bibinfo {author} {\bibfnamefont {S.}~\bibnamefont
  {Uchino}},\ }\href {\doibase 10.1088/1367-2630/aad1e7} {\bibfield  {journal}
  {\bibinfo  {journal} {New J. Phys.}\ }\textbf {\bibinfo {volume} {20}},\
  \bibinfo {pages} {073048} (\bibinfo {year} {2018})}\BibitemShut {NoStop}%
\bibitem [{\citenamefont {Mulkerin}\ \emph {et~al.}()\citenamefont {Mulkerin},
  \citenamefont {Liu},\ and\ \citenamefont {Hu}}]{Mulkerin2018}%
  \BibitemOpen
  \bibfield  {author} {\bibinfo {author} {\bibfnamefont {B.~C.}\ \bibnamefont
  {Mulkerin}}, \bibinfo {author} {\bibfnamefont {X.-J.}\ \bibnamefont {Liu}}, \
  and\ \bibinfo {author} {\bibfnamefont {H.}~\bibnamefont {Hu}},\ }\href
  {http://arxiv.org/abs/1808.07671} {\bibinfo  {journal} {arXiv:1808.07671}
  }\BibitemShut {NoStop}%
\bibitem [{Note2()}]{Note2}%
  \BibitemOpen
\bibfield  {journal} {  }\bibinfo {note} {Recently, a preprint appeared in
  which the authors, motivated by our work, found similar sudden shifts of the
  peak rf transfer. See H. Tajima and S. Uchino, arXiv:1812.05889.}\BibitemShut
  {Stop}%
\bibitem [{\citenamefont {Enss}\ \emph {et~al.}(2011)\citenamefont {Enss},
  \citenamefont {Haussmann},\ and\ \citenamefont {Zwerger}}]{Enss2011}%
  \BibitemOpen
  \bibfield  {author} {\bibinfo {author} {\bibfnamefont {T.}~\bibnamefont
  {Enss}}, \bibinfo {author} {\bibfnamefont {R.}~\bibnamefont {Haussmann}}, \
  and\ \bibinfo {author} {\bibfnamefont {W.}~\bibnamefont {Zwerger}},\ }\href
  {\doibase 10.1016/j.aop.2010.10.002} {\bibfield  {journal} {\bibinfo
  {journal} {Ann. Phys. (Amsterdam)}\ }\textbf {\bibinfo {volume} {326}},\
  \bibinfo {pages} {770} (\bibinfo {year} {2011})}\BibitemShut {NoStop}%
\bibitem [{\citenamefont {Sun}\ and\ \citenamefont {Leyronas}(2015)}]{Sun2015}%
  \BibitemOpen
  \bibfield  {author} {\bibinfo {author} {\bibfnamefont {M.}~\bibnamefont
  {Sun}}\ and\ \bibinfo {author} {\bibfnamefont {X.}~\bibnamefont {Leyronas}},\
  }\href {\doibase 10.1103/PhysRevA.92.053611} {\bibfield  {journal} {\bibinfo
  {journal} {Phys. Rev. A}\ }\textbf {\bibinfo {volume} {92}},\ \bibinfo
  {pages} {053611} (\bibinfo {year} {2015})}\BibitemShut {NoStop}%
\bibitem [{\citenamefont {Ho}\ and\ \citenamefont {Mueller}(2004)}]{Ho2004a}%
  \BibitemOpen
  \bibfield  {author} {\bibinfo {author} {\bibfnamefont {T.-L.}\ \bibnamefont
  {Ho}}\ and\ \bibinfo {author} {\bibfnamefont {E.~J.}\ \bibnamefont
  {Mueller}},\ }\href {\doibase 10.1103/PhysRevLett.92.160404} {\bibfield
  {journal} {\bibinfo  {journal} {Phys. Rev. Lett.}\ }\textbf {\bibinfo
  {volume} {92}},\ \bibinfo {pages} {160404} (\bibinfo {year}
  {2004})}\BibitemShut {NoStop}%
\bibitem [{\citenamefont {Pieri}\ \emph {et~al.}(2009)\citenamefont {Pieri},
  \citenamefont {Perali},\ and\ \citenamefont {Strinati}}]{Pieri2009}%
  \BibitemOpen
  \bibfield  {author} {\bibinfo {author} {\bibfnamefont {P.}~\bibnamefont
  {Pieri}}, \bibinfo {author} {\bibfnamefont {A.}~\bibnamefont {Perali}}, \
  and\ \bibinfo {author} {\bibfnamefont {G.~C.}\ \bibnamefont {Strinati}},\
  }\href {\doibase 10.1038/nphys1345} {\bibfield  {journal} {\bibinfo
  {journal} {Nat. Phys.}\ }\textbf {\bibinfo {volume} {5}},\ \bibinfo {pages}
  {736} (\bibinfo {year} {2009})}\BibitemShut {NoStop}%
\bibitem [{\citenamefont {Stewart}\ \emph {et~al.}(2010)\citenamefont
  {Stewart}, \citenamefont {Gaebler}, \citenamefont {Drake},\ and\
  \citenamefont {Jin}}]{Stewart2010}%
  \BibitemOpen
  \bibfield  {author} {\bibinfo {author} {\bibfnamefont {J.~T.}\ \bibnamefont
  {Stewart}}, \bibinfo {author} {\bibfnamefont {J.~P.}\ \bibnamefont
  {Gaebler}}, \bibinfo {author} {\bibfnamefont {T.~E.}\ \bibnamefont {Drake}},
  \ and\ \bibinfo {author} {\bibfnamefont {D.~S.}\ \bibnamefont {Jin}},\ }\href
  {\doibase 10.1103/PhysRevLett.104.235301} {\bibfield  {journal} {\bibinfo
  {journal} {Phys. Rev. Lett.}\ }\textbf {\bibinfo {volume} {104}},\ \bibinfo
  {pages} {235301} (\bibinfo {year} {2010})}\BibitemShut {NoStop}%
\bibitem [{\citenamefont {Yan}\ and\ \citenamefont {Blume}(2013)}]{Yan2013}%
  \BibitemOpen
  \bibfield  {author} {\bibinfo {author} {\bibfnamefont {Y.}~\bibnamefont
  {Yan}}\ and\ \bibinfo {author} {\bibfnamefont {D.}~\bibnamefont {Blume}},\
  }\href {\doibase 10.1103/PhysRevA.88.023616} {\bibfield  {journal} {\bibinfo
  {journal} {Phys. Rev. A}\ }\textbf {\bibinfo {volume} {88}},\ \bibinfo
  {pages} {023616} (\bibinfo {year} {2013})}\BibitemShut {NoStop}%
\bibitem [{\citenamefont {Trefzger}\ and\ \citenamefont
  {Castin}(2013)}]{Trefzger2013}%
  \BibitemOpen
  \bibfield  {author} {\bibinfo {author} {\bibfnamefont {C.}~\bibnamefont
  {Trefzger}}\ and\ \bibinfo {author} {\bibfnamefont {Y.}~\bibnamefont
  {Castin}},\ }\href {\doibase 10.1209/0295-5075/101/30006} {\bibfield
  {journal} {\bibinfo  {journal} {Europhys. Lett.}\ }\textbf {\bibinfo {volume}
  {101}},\ \bibinfo {pages} {30006} (\bibinfo {year} {2013})}\BibitemShut
  {NoStop}%
\bibitem [{\citenamefont {Liu}\ and\ \citenamefont {Hu}(2010)}]{Liu2010b}%
  \BibitemOpen
  \bibfield  {author} {\bibinfo {author} {\bibfnamefont {X.-J.}\ \bibnamefont
  {Liu}}\ and\ \bibinfo {author} {\bibfnamefont {H.}~\bibnamefont {Hu}},\
  }\href {\doibase 10.1103/PhysRevA.82.043626} {\bibfield  {journal} {\bibinfo
  {journal} {Phys. Rev. A}\ }\textbf {\bibinfo {volume} {82}},\ \bibinfo
  {pages} {043626} (\bibinfo {year} {2010})}\BibitemShut {NoStop}%
\bibitem [{\citenamefont {Mora}\ and\ \citenamefont {Chevy}(2010)}]{Mora2010}%
  \BibitemOpen
  \bibfield  {author} {\bibinfo {author} {\bibfnamefont {C.}~\bibnamefont
  {Mora}}\ and\ \bibinfo {author} {\bibfnamefont {F.}~\bibnamefont {Chevy}},\
  }\href {\doibase 10.1103/PhysRevLett.104.230402} {\bibfield  {journal}
  {\bibinfo  {journal} {Phys. Rev. Lett.}\ }\textbf {\bibinfo {volume} {104}},\
  \bibinfo {pages} {230402} (\bibinfo {year} {2010})}\BibitemShut {NoStop}%
\bibitem [{\citenamefont {Punk}\ \emph {et~al.}(2009)\citenamefont {Punk},
  \citenamefont {Dumitrescu},\ and\ \citenamefont {Zwerger}}]{Punk2009}%
  \BibitemOpen
  \bibfield  {author} {\bibinfo {author} {\bibfnamefont {M.}~\bibnamefont
  {Punk}}, \bibinfo {author} {\bibfnamefont {P.~T.}\ \bibnamefont
  {Dumitrescu}}, \ and\ \bibinfo {author} {\bibfnamefont {W.}~\bibnamefont
  {Zwerger}},\ }\href {\doibase 10.1103/PhysRevA.80.053605} {\bibfield
  {journal} {\bibinfo  {journal} {Phys. Rev. A}\ }\textbf {\bibinfo {volume}
  {80}},\ \bibinfo {pages} {053605} (\bibinfo {year} {2009})}\BibitemShut
  {NoStop}%
\bibitem [{\citenamefont {Schmidt}\ and\ \citenamefont
  {Enss}(2011)}]{Schmidt2011}%
  \BibitemOpen
  \bibfield  {author} {\bibinfo {author} {\bibfnamefont {R.}~\bibnamefont
  {Schmidt}}\ and\ \bibinfo {author} {\bibfnamefont {T.}~\bibnamefont {Enss}},\
  }\href {\doibase 10.1103/PhysRevA.83.063620} {\bibfield  {journal} {\bibinfo
  {journal} {Phys. Rev. A}\ }\textbf {\bibinfo {volume} {83}},\ \bibinfo
  {pages} {063620} (\bibinfo {year} {2011})}\BibitemShut {NoStop}%
\bibitem [{\citenamefont {Schmidt}\ \emph {et~al.}(2018)\citenamefont
  {Schmidt}, \citenamefont {Knap}, \citenamefont {Ivanov}, \citenamefont {You},
  \citenamefont {Cetina},\ and\ \citenamefont {Demler}}]{Schmidt2018}%
  \BibitemOpen
  \bibfield  {author} {\bibinfo {author} {\bibfnamefont {R.}~\bibnamefont
  {Schmidt}}, \bibinfo {author} {\bibfnamefont {M.}~\bibnamefont {Knap}},
  \bibinfo {author} {\bibfnamefont {D.~A.}\ \bibnamefont {Ivanov}}, \bibinfo
  {author} {\bibfnamefont {J.-S.}\ \bibnamefont {You}}, \bibinfo {author}
  {\bibfnamefont {M.}~\bibnamefont {Cetina}}, \ and\ \bibinfo {author}
  {\bibfnamefont {E.}~\bibnamefont {Demler}},\ }\href {\doibase
  10.1088/1361-6633/aa9593} {\bibfield  {journal} {\bibinfo  {journal} {Rep.
  Prog. Phys.}\ }\textbf {\bibinfo {volume} {81}},\ \bibinfo {pages} {024401}
  (\bibinfo {year} {2018})}\BibitemShut {NoStop}%
\end{thebibliography}
\end{document}


\pagebreak
\onecolumngrid
\begin{center}
\large{\textbf{Supplemental Material:\\Boiling a Unitary Fermi Liquid}}
\end{center}
\twocolumngrid
\setcounter{equation}{0}
\setcounter{figure}{0}
\setcounter{table}{0}
\setcounter{page}{1}
\makeatletter
\renewcommand{\thepage}{S\arabic{page}} 
\renewcommand{\theequation}{S\arabic{equation}}
\renewcommand{\thefigure}{S\arabic{figure}}
\renewcommand{\bibnumfmt}[1]{[S#1]}
\renewcommand{\citenumfont}[1]{S#1}

\author{Zhenjie Yan}
\affiliation{\affcua}
\author{Parth B. Patel}
\affiliation{\affcua}
\author{Biswaroop Mukherjee}
\affiliation{\affcua}

\author{Richard J. Fletcher}
\affiliation{\affcua}

\author{Julian Struck}
\affiliation{\affcua}
\affiliation{\affens}

\author{Martin W. Zwierlein}
\affiliation{\affcua}

\maketitle


\section{State Preparation}

The three hyperfine states $\ket{\downarrow}=\ket{1}$, $\ket{\mathrm{f}}=\ket{2}$ and $\ket{\uparrow}=\ket{3}$ are adiabatically connected to the respective states $\ket{F=\frac{1}{2},m_{F}=\frac{1}{2}}$, $\ket{\frac{1}{2},-\frac{1}{2}}$ and $\ket{\frac{3}{2},-\frac{3}{2}}$ at vanishing magnetic field. As in previous works at MIT \cite{Schunck2008}, we start from a degenerate and fully spin-polarized gas in state $\ket{1}$ in an optical dipole trap, and prepare the spin-imbalanced mixture through two consecutive Landau-Zener sweeps at a magnetic field of $B=569\,\mathrm{G}$, where the interactions between all three states are weak.
The magnetic field is then ramped up within $1\,\mathrm{ms}$ to the $\ket{1}$-$\ket{3}$ Feshbach resonance at $B=690\,\mathrm{G}$, where the gas is loaded into the uniform potential and cooled through forced evaporation over 2 seconds \cite{Mukherjee2017b}.

\begin{figure}[!b]
\centering
\includegraphics[width=8.6cm]{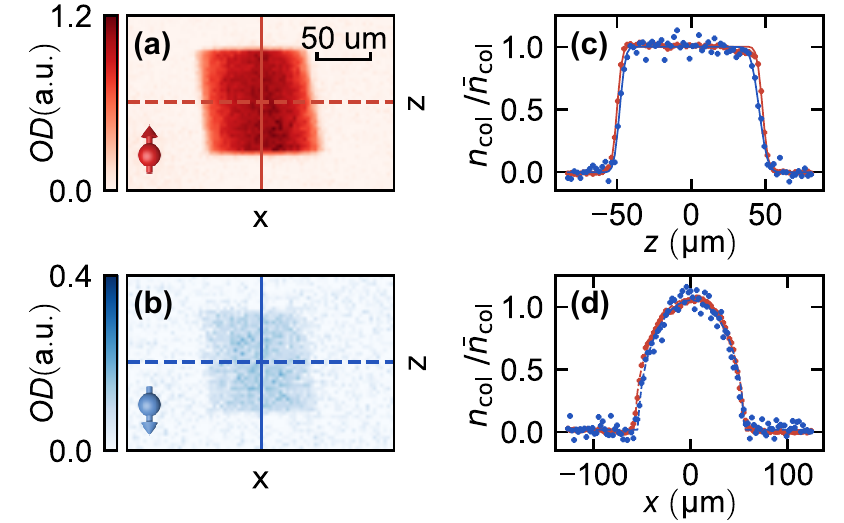}
\caption{
Spin-imbalanced Fermi gas in the cylinder-shaped optical trap. Here, $E_{\mathrm{F}\uparrow}/h = 10.3 \pm 0.1 \, \mathrm{kHz}$, $T /T_{\mathrm{F}\uparrow}=0.74\pm0.01$ and $n_{\downarrow}/n_{\uparrow}=0.1$. 
\textit{In situ} absorption images of the majority (a) and minority (b) component along the radial direction of the cylinder.
The column density along the axial (c) and radial (d) direction. The red (blue) points correspond to the majority (minority) component. The solid lines indicate the fits with Eqs. \eqref{eq:RadialDensity} and \eqref{eq:AxialDensity}.
} 
\label{fig:SM1}
\end{figure}

\section{Uniform Cylinder Trap}

Uniform trapping potentials offer a crucial advantage for non-local rf spectroscopy. For non-uniform traps, the Fermi energy and the impurity concentration vary spatially, leading to an artificial broadening of the rf spectra.

For the rf measurements we load the gas into a cylinder-shaped uniform optical trap \cite{Mukherjee2017b} [Fig.\ref{fig:SM1} (a) and (b)]. Forced evaporation in this trap leads to temperatures of $T \simeq 0.05\,T_{\mathrm{F}\uparrow}$ and majority Fermi energies of $E_{\mathrm{F}\uparrow}/h \sim 10\, \mathrm{kHz}$. Instead of adjusting the temperature of the gas by evaporation, we introduce an additional heating step by periodically modulating the cylindrical trapping potential by $20\%$ at $1\,\mathrm{kHz}$ up to $4.5\,\mathrm{s}$. We have found that this additional step reduces the spread in Fermi energies for different temperatures compared to a control through the evaporation.

We directly probe the variation in impurity concentration in the cloud from in-situ measurements of the cloud profiles [Fig.\ref{fig:SM1} (c) and (d)]. We find that the ratio of the effective volume of the minority component and the majority component is $V_{\downarrow}/V_{\uparrow} > 0.8$. More significantly for our purposes, well over 90\% of the minority cloud is in a region of constant majority density at all temperatures. The radius $R_{\uparrow,\downarrow}$ and length $L_{\uparrow,\downarrow}$ of the gas have been determined by fitting the column integrated density profiles of the minority and majority component. The fitting functions along the radial (x) and the axial (z) direction are given by:
\begin{align}
	n_{\mathrm{col}}(x)=\, &\bar{n}_{\mathrm{col}} \sqrt{R^2-x^2} ~ , \label{eq:RadialDensity}
	\\
	n_{\mathrm{col}}(z) = \, &\bar{n}_{\mathrm{col}} \bigg( \mathrm{erf} \left( \frac{L/2+z}{\sqrt{2}\sigma_1} \right) \nonumber
	\\
	&+\mathrm{erf} \left( \frac{L/2-z}{\sqrt{2}\sigma_2} \right) \bigg)/2 ~. \label{eq:AxialDensity}
\end{align}
Here, $\mathrm{erf}(z)$ denotes the error function and $\sigma_{1,2}$ the effective widths of the potentials walls along the axial direction.

\section{Thermometry of the homogeneous Unitary Fermi Gas}

Standard thermometry methods for ultracold atoms rely on the low-fugacity tails in non-uniform traps. In addition, for weakly or noninteracting gases the momentum distribution, measured in time-of-flight, can be used to infer the temperature. However, in the case of strongly interacting homogeneous gases none of these methods is directly applicable. Here we describe our thermometry method to obtain the temperature of the unitary Fermi gas in a uniform trap. First, we determine the energy of the gas in the uniform box potential. This energy then yields the temperature using the equation of state of the imbalanced gas.

\subsection{Energy Measurement}

\begin{figure}
\centering
\includegraphics[width=8.6cm]{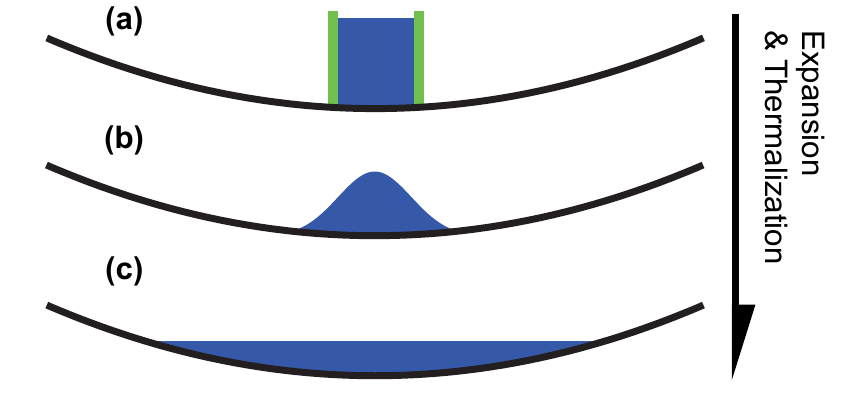}
\caption{Isoenergetic expansion from the uniform to harmonic trap. (a) Initially the gas is trapped in a quasi-uniform potential, (b) then the end caps of the trap are instantaneously removed, allowing the gas to expand into a harmonic trapping potential along the axial direction. (c) After a $2\,\mathrm{s}$ hold time the gas is in thermal equilibrium and the energy is determined using a one-dimensional virial theorem.} 
\label{fig:SM2}
\end{figure}

The total energy of the homogeneous gas is measured by an isoenergetic expansion into a hybrid cylindrical trap that features a harmonic potential along the axial direction ($\omega_{z}=2\pi \cdot 23\,\mathrm{Hz}$) and is uniform in the radial direction \cite{Mukherjee2017b}. Figure \ref{fig:SM2} shows the isoenergetic transfer between the two trapping potentials. For the transfer the two endcaps of the cylindrical trap are instantaneously removed and therefore no work is performed on the atoms. Subsequently the gas expands isoenergetically along the axial direction. Note that the underlying harmonic potential along the axial direction is always present, even for the experiments in the uniform trap. However, the potential variation due to the harmonic potential is only a few percent of the Fermi energy \cite{Mukherjee2017b}. After the removal of the endcaps we wait for $2\,\mathrm{s}$ to equilibrate the gas. The total energy of the gas in the hybrid cylindrical trap can be determined from the \emph{in situ} density profiles using a one-dimensional version of the virial theorem \cite{Thomas2005}. 

The total energy of the system is given by sum of the internal and potential energy
\begin{equation}
E = \int \mathrm{d}^3 r \big(\epsilon(r) + n(r) \, U(r) \big),
\label{eqn:GeneralTotalEnergy}
\end{equation}
with the internal energy density $\epsilon(r)$, the total density $n(r)=n_{\downarrow}(r)+n_{\uparrow}(r)$ and the potential energy $U(r)$. The hybrid trap is uniform along the radial direction and we can express Eq. \eqref{eqn:GeneralTotalEnergy} with the cylindrical cross-section $A_{\mathrm{cyl}}$ as 
\begin{equation}
E = A_{\mathrm{cyl}} \int \mathrm{d}z \big(\epsilon(z) + n(z) \, U(z) \big),
\label{eqn:1DTotalEnergy}
\end{equation}
where we have defined $\epsilon(z) \equiv \epsilon(x=0,y=0,z)$, $n(z) \equiv n(x=0,y=0,z)$ and $U(z) \equiv U(x=0,y=0,z)$.
For the unitary Fermi gas the internal energy is directly related to the pressure $\epsilon(z)= 3/2 \, P(z)$ \cite{Ho2004}. Combined with a partial integration of the first term in Eq. \eqref{eqn:1DTotalEnergy} this leads to
\begin{equation}
E = - \frac{3}{2}\,A_{\mathrm{cyl}} \int z \frac{\partial{P(z)}}{\partial{z}}\,\mathrm{d}z + A_{\mathrm{cyl}}  \int n(z) U(z) \,\mathrm{d}z \, , 
\label{eqn:1DTotalEnergyIntermediateResult1}
\end{equation}
where we have assumed that the pressure $P(z)$ is an even function of $z$. With the equation for hydrostatic equilibrium (Gibbs-Duhem equation at constant scattering length and temperature) $\mathrm{d}P=n\,\mathrm{d}\mu$ and the local density approximation  $\mathrm{d}\mu = -\mathrm{d}U$, we obtain the one-dimensional virial theorem
\begin{align}
E &= \frac{3}{2}\,A_{\mathrm{cyl}} \int z\,n(z) \frac{\partial{U(z)}}{\partial{z}}\,\mathrm{d}z + A_{\mathrm{cyl}} \int n(z) U(z) \,\mathrm{d}z \, , \nonumber
\\ 
&=  2\, A_{\mathrm{cyl}} \, m \, \omega_{\mathrm{z}}^2 \, \int n(z) \, z^2 \, \mathrm{d}z \, .
\label{eqn:1Dvirial}
\end{align}
In the second step of Eq. \eqref{eqn:1Dvirial} we have used the explicit expression for the harmonic potential $U(z) = m \, \omega_{\mathrm{z}}^2 \, z^2 \, /2$, with the trapping frequency $\omega_{\mathrm{z}}$ and mass $m$.

\subsection{Equation of State of the \\ Spin-Imbalanced Unitary Fermi Gas}

\begin{figure}
\centering
\includegraphics[width=8.6cm]{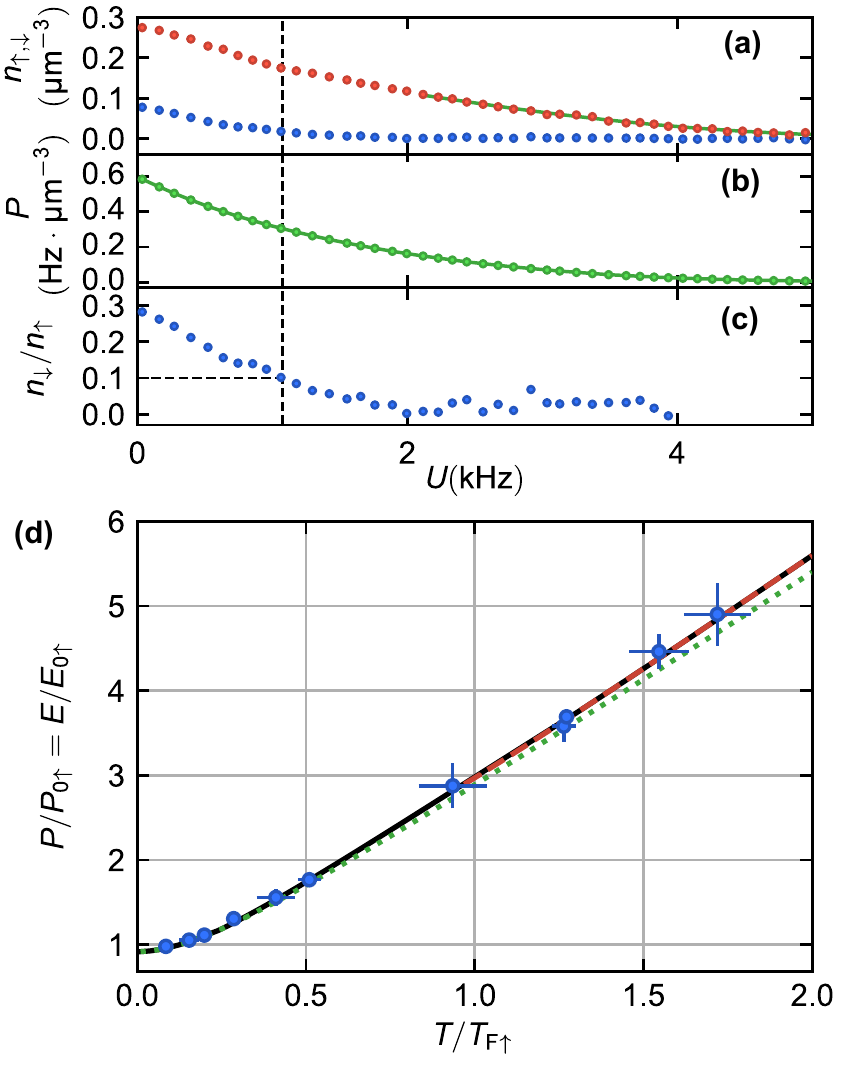}
\caption{
Pressure of the spin-imbalanced unitary Fermi gas.
(a) An example of \emph{in situ} density profiles of the majority (red) and minority (blue) component in the hybrid trap. The temperature of $T=31\pm2\,\mathrm{nK}$ is determined from a fit of the spin-polarized tail of the majority component with the equation of state of the noninteracting Fermi gas (green solid line). The majority Fermi energy in the center of the trap ($U=0$) is $E_{\mathrm{F}\uparrow}/h=5.4\pm0.2 \, \mathrm{kHz}$. (b) The local pressure of the gas, obtained from integrating the density profiles. (c) The local minority concentration. The relevant values for a minority concentration of $n_{\downarrow}/n_{\uparrow}=0.1$ are indicated with the dashed line. (d) The equation of state of the gas at fixed impurity concentration of $n_{\downarrow}/n_{\uparrow}=0.1$.
Our measurement is compared with the Fermi liquid ansatz (green dotted line) and 3rd order Virial expansion (red dashed line).
An interpolation function (black solid line) is used to connect the Fermi liquid ansatz and Virial expansion.} 
\label{fig:SM3}
\end{figure}

The conversion of the total energy into the corresponding temperature requires knowledge of the finite-temperature equation of state of the system. For this purpose, we have measured the finite temperature pressure equation of state of the spin-imbalanced unitary Fermi gas. As mentioned previously, at unitarity pressure and energy are directly related through $E = 3/2 \, P\,V$, with V as the volume \cite{Ho2004}. The zero-temperature equation of state of the spin-imbalanced Fermi gas has been measured previously, confirming Fermi-Liquid behaviour beyond the Chandrasekhar-Clogston limit in the normal fluid phase \cite{Shin2008,Shin2008a,Nascimbene2010,Navon2010,Nascimbene2011}. We determine the pressure of the gas from \emph{in situ} density profiles in the hybrid trap \cite{Mukherjee2017b} and a precise knowledge of the harmonic trapping potential $U$ along the axial direction of the trap [Fig. \ref{fig:SM3}(a)]. From the equation for hydrostatic equilibrium and the local density approximation follows for the pressure $P(U)=\int_{U}^{\infty} n(U) \mathrm{d} U$ \cite{Ku2012} [Fig. \ref{fig:SM3}(b)]. The impurity concentration is varying in this non-uniform trapping potential [Fig. \ref{fig:SM3}(c)]. To fill the three-dimensional parameter space $P(n,n_{\downarrow}/n_{\uparrow},T)$, we measure density profiles for varying initial evaporation parameters. The temperature is obtained from third-order virial expansion fits \cite{Liu2010b} to the low-fugacity tails of the gas or fits with the equation of state of the noninteracting Fermi gas, in case of spin-polarized tails.

Figure \ref{fig:SM3}(d) shows the equation of state at a fixed impurity concentration of $n_{\downarrow}/n_{\uparrow}=0.1$. For the unitary Fermi gas, the normalized pressure and energy are identical: $P/P_{0\uparrow}=E/E_{0\uparrow}$, with the ground state energy $E_{0\uparrow}=3/5\,N_{\uparrow}E_{\mathrm{F}\uparrow}$ and pressure $P_{0\uparrow} =2/5\,n_{\uparrow}E_{\mathrm{F}\uparrow}$ of the majority atoms. At low temperatures, the normalized pressure is in agreement with the Fermi liquid pressure ansatz [Eq. (3)], while at higher temperatures it is in agreement with the virial expansion. We use an interpolation function that connects the high and low temperature regime to be able to determine the temperature for arbitrary pressure (energy) values [Fig. \ref{fig:SM3}(d)].

\section{Radio-frequency Spectroscopy}

In this section, we discuss the detection scheme used for rf spectroscopy, the determination of the transfer rate in the linear response regime and the definition of the error bars in Fig.1 (b) and (c).

\subsection{Detection}

For the measurement of the normalized transfer $I(\omega)$, the atom numbers in all three spin states $\ket{1}$, $\ket{2}$ and $\ket{3}$ after the rf pulse need to be determined. For this purpose we have implemented a triple absorption imaging scheme that allows us to detect all three spin components in a single experimental run. The amount of transferred atoms $N_{\mathrm{f}}$ is determined from an absorption image along the axial direction of the cylinder trap recorded on an EMCCD camera. After $3 \,\mathrm{ms}$, the spin components $\ket{1}$ and $\ket{3}$ are subsequently imaged along the radial direction of the cylinder trap and recorded on a fast SCMOS camera with a time delay of $\mathrm{10\,\mu s}$ between the two states.

\subsection{Linear Response}

Linear response is an essential requirement for the applicability of Eqs. (1) and (2), linking the rf spectra to the spectral function and the contact. The linear transfer rate can be determined by measuring the transfer fraction $N_{\mathrm{f}}/N_\downarrow$ as a function of the rf pulse duration [Fig. \ref{fig:SM4}]. For the contact data we fit the transfer rate with the exponential function $N_\mathrm{f} / N_\downarrow=\Gamma\tau(1-\mathrm{exp}(-t/\tau))$ to account for saturation effects (time scale $\tau$) and use the initial slope at short times to derive the linear transfer rate $\Gamma$ [Fig. \ref{fig:SM4}].

\subsection{Error Bars in Figure 1}

\begin{figure}
\centering
\includegraphics[width=8.6cm]{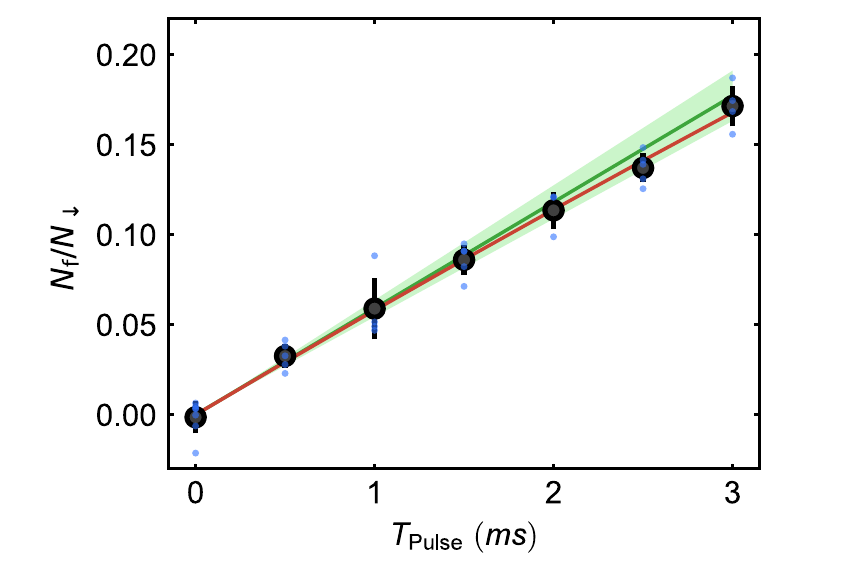}
\caption{Time resolved rf response of the minority component. 
Here, the majority Fermi energy is $E_{\mathrm{F}\uparrow}/h=9.9\pm 0.2 \, \mathrm{kHz}$, the impurity concentration $n_{\downarrow}/n_{\uparrow}=0.11\pm0.02$, the rf detuning $\omega=2\pi \times 60\, \mathrm{kHz}$ and the temperature $T/T_{\mathrm{F}\uparrow}=0.73\pm0.02$. 
The blue datapoints correspond to the measured rf transfer for several experimental runs and the black circles are the averaged data points. 
The transferred fraction is fitted with an exponential function (red solid line). The green solid line (shaded area) shows the initial linear slope $\Gamma$ (standard error) of the fit.} 
\label{fig:SM4}
\end{figure}

Figure 1 shows the peak position and full width at half maximum (FWHM) at different temperatures. To determine the experimental uncertainty for these observables, we use the standard error of the measured transferred fraction without any rf pulse (see $T_{\mathrm{Pulse}}=0$ in Fig. \ref{fig:SM4}). The errors for the peak position and FWHM are then inferred from the frequency range of data points that lie within the standard error of the amplitude of the maximum and half maxima respectively.

%